\documentclass{article}
\usepackage{fullpage}
\usepackage[utf8]{inputenc}

\title{Learning joint and individual structure in network data with covariates}
\author{Carson James \thanks{Department of Statistics, Texas A\&M University}, Dongbang Yuan \thanks{Meta}, Irina Gaynanova \thanks{Department of Biostatistics, University of Michigan}, Jes\'us Arroyo \thanks{Department of Statistics, Texas A\&M University}
\thanks{Corresponding author (Email: \href{jarroyo@tamu.edu}{jarroyo@tamu.edu}).}}

\usepackage{amsmath}
\usepackage{amssymb}
\usepackage{amsthm}
\usepackage{comment}
\usepackage{xcolor}
\usepackage{hyperref}
\usepackage{graphicx}
\usepackage{algorithmicx}
\usepackage{algorithm}
\usepackage{algpseudocode}
\graphicspath{ {./Figures/Trade_Data_Analysis/} }
\usepackage{natbib}
\usepackage{subcaption}
\usepackage{wrapfig}
\usepackage{tikz-cd} 

\usepackage{amsmath}
\usepackage{amsfonts}
\usepackage{amssymb}
\usepackage{amsthm}
\usepackage{dsfont}
\usepackage{subcaption}
\usepackage[shortlabels]{enumitem}
\usepackage{fullpage}

\usepackage{algorithm}
\usepackage{algpseudocode}

\usepackage{natbib}

\newtheorem{thm}{Theorem}[section]
\newtheorem{prop}[thm]{Proposition}
\newtheorem{lem}[thm]{Lemma}
\newtheorem{cor}[thm]{Corollary}

\theoremstyle{definition}
\newtheorem{defn}{Definition}[section]

\newtheorem{asmp}{Assumption}

\newcommand{\wt}[1]{\widetilde{#1}}

\def\R{\mathbb{R}}
\def\E{\mathbb{E}} 

\def\R{\mathbb{R}}
\def\E{\mathbb{E}}

\def\N{\mathbb{N}}
\def\O{\mathbb{O}}

\DeclareMathOperator*{\Var}{Var}

\DeclareMathOperator*{\minimize}{minimize}

\newcommand{\bpm}{\begin{pmatrix}}
\newcommand{\epm}{\end{pmatrix}}
\newcommand{\ee}{\end{equation}}
\newcommand{\bes}{\begin{equation*}}
\newcommand{\ees}{\end{equation*}}

\def\boxit#1{\vbox{\hrule\hbox{\vrule\kern6pt  \vbox{\kern6pt#1\kern6pt}\kern6pt\vrule}\hrule}}

\def\R{\mathbb{R}}

\newcommand{\lam}{\lambda}
\newcommand{\sig}{\sigma}
\newcommand{\del}{\delta}

\newcommand{\ep}{\epsilon}
\newcommand{\al}{\alpha}
\newcommand{\om}{\omega}
\newcommand{\Om}{\Omega}

\newcommand{\Lam}{\Lambda}
\newcommand{\Sig}{\Sigma}

\newcommand{\Gam}{\Gamma}

\newcommand{\MC}{\mathcal{C}}
\newcommand{\MM}{\mathcal{M}}
\newcommand{\MR}{\mathcal{R}}
\newcommand{\MP}{\mathcal{P}}
\newcommand{\MQ}{\mathcal{Q}}

\newcommand{\MU}{\mathcal{U}}
\newcommand{\MV}{\mathcal{V}}

\newcommand{\wh}[1]{\widehat{#1}}
\newcommand{\tbf}[1]{\textbf{#1}}

\renewcommand{\l}{\langle}
\renewcommand{\r}{\rangle}

\DeclareMathOperator{\rnk}{rank}

\DeclareMathOperator{\eig}{EIG}
\DeclareMathOperator{\sv}{SV}

\newcommand{\labthm}[1]{\label{thm:#1}}
\newcommand{\refthm}[1]{Theorem \ref{thm:#1}}

\newcommand{\labprop}[1]{\label{prop:#1}}
\newcommand{\refprop}[1]{Proposition \ref{prop:#1}}

\newcommand{\labcor}[1]{\label{cor:#1}}
\newcommand{\refcor}[1]{Corollary \ref{cor:#1}}

\newcommand{\lablem}[1]{\label{lemma:#1}}
\newcommand{\reflem}[1]{Lemma \ref{lemma:#1}}

\newcommand{\labeq}[1]{\label{eq:#1}}
\newcommand{\refeq}[1]{Equation (\ref{eq:#1})}

\newcommand{\labfig}[1]{\label{fig:#1}}
\newcommand{\reffig}[1]{Figure \ref{fig:#1}}

\newcommand{\labalg}[1]{\label{alg:#1}}
\newcommand{\refalg}[1]{Algorithm \ref{alg:#1}}

\newcommand{\labdef}[1]{\label{defn:#1}}
\newcommand{\refdef}[1]{Definition \ref{defn:#1}}

\newcommand{\labtable}[1]{\label{table:#1}}
\newcommand{\reftable}[1]{Table \ref{table:#1}}

\newcommand{\labproof}[1]{\label{proof:#1}}

\newcommand{\labasmp}[1]{\label{asmp:#1}}
\newcommand{\refasmp}[1]{Assumption \ref{asmp:#1}}

\algnewcommand\algorithmicinput{\textbf{Input:}}
\algnewcommand\Input{\item[\algorithmicinput]}
\algnewcommand\algorithmiciterate{\textbf{Iterate:}}
\algnewcommand\Iterate{\item[\algorithmiciterate]}
\algnewcommand\algorithmicinitialize{\textbf{Initialize:}}
\algnewcommand\Initialize{\item[\algorithmicinitialize]}
\algnewcommand\algorithmicoutput{\textbf{Output:}}
\algnewcommand\Output{\item[\algorithmicoutput]}
\algnewcommand\RETURN{\State \algorithmicreturn}%

\begin{document}

	\maketitle
	
	\begin{abstract}
		Datasets consisting of a network and covariates associated with its vertices have become ubiquitous. One problem pertaining to this type of data is to identify information unique to the network, information unique to the vertex covariates and information that is shared between the network and the vertex covariates. Existing techniques for network data and vertex coviariates focus on capturing structure that is shared but are usually not able to differentiate structure that is unique to each dataset. This work formulates a low-rank model that simultaneously captures joint and individual information in network data with vertex covariates. A two-step estimation procedure is proposed,  composed of an efficient spectral method followed by a refinement optimization step. Theoretically, we show that the spectral method is able to consistently recover the joint and individual components under a general signal-plus-noise model.  
  Simulations and real data examples demonstrate the ability of the methods to recover accurate and interpretable components. In particular, the application of the methodology to a food trade network between countries with economic, developmental and geographical country-level indicators as covariates yields joint and individual factors that explain the trading patterns.
	\end{abstract}
	
	\section{Introduction}
	
	Network data is ubiquitous in many disciplines and application domains, including computer science, statistics, biology, and physics. These data, encoding relationships between units represented as nodes, are often accompanied by additional information about the nodes, usually referred to as node covariates, attributes, or metadata \citep{Newman:2016:annotated_networks,Liu:2019:statistical_learning_networks_features,Chunaev:2020:node_attributed_networks_survey}. In these situations, a common goal is to understand the associations between the network connectivity and the node covariates. In our example, we consider international food commodity trade data represented as a network, where the nodes correspond to different countries and edge weights encode food commodity trade volumes between corresponding countries. The covariates at each node consist of economic and geographic information for each country, such as gross domestic product (GDP) per capita, birth rate and region. We wish to exploit that both datasets contain information about the nodes in order to better understand the structure of the network, node covariates and their relationship. Specifically, we seek to understand how economic and geographic factors explain the observed trade between countries, and identify additional information in the network that cannot be explained solely by these variables. 
 
	There has been substantial work that incorporates network and node covariate information. Some examples include methods that use node covariates to improve community detection \citep{Binkiewicz:2017:casc,Huang:2023:pcabm}, dimensionality reduction \citep{Zhao:2022:dim_reduct_net_covar}, regression with network information \citep{Li:2019:predict_network_linked_data} and mixed effect models for network edges \citep{Hoff:2005:bilinear_mixed_effects}. These existing methods aim to exploit the common information in the network and its node covariates but do not address whether the covariates contain information that is not captured by the network or vice-versa. Such methods may also discard information contained in only network or covariate data.

    When integrating multi-view datasets with matched units, a common task is to identify joint and individual structures across these datasets, for example, by partitioning the data into components that either capture shared information or information that is unique to each individual dataset \citep{Lock:2013:jive, Feng:2018:ajive}. Existing work in this area has focused on multi-view data matrices. In constrast, network data provides information about interactions or relationships between units rather than unit-level data, making integration of such data more challenging. Recent studies have explored integration problems in the context of multiple network datasets with matched units \citep{MacDonald:2022:multiplex_latent_space}. To the best of our knowledge, the task of integrating network data with matrix data (representing edge or node covariates) with the goal of identifying joint and individual signals across these two distinct sources has not been explored in prior works.
	
	In this work, we study the problem of integrating network data with node covariates making the following contributions: 
 \begin{itemize}
     \item We propose a new unified model that partitions the information among node connectivity and node covariate data into joint and individual components. These node components can be interpreted as node embeddings that capture either joint or individual structure, and we discuss the connections between these embeddings and latent space models \citep{Hoff:2002:latent_social}.
     \item We develop an estimation procedure based on a computationally efficient spectral method followed by an optimization-based refining step. We derive a fast optimization algorithm for the latter based on closed-form block-coordinate descent updates. 
     \item Under a general signal-plus-noise error model, we derive an explicit characterization of the error rate for the joint and individual components obtained from the spectral method and prove consistency of the spectral estimates under reasonable restrictions on the model parameters.
     \item We showcase that the proposed method effectively captures joint and individual information via simulations and the analysis of an international food commodity trade dataset paired with econometric data. We find that the joint components for the network and node covariates capture joint trading structure related to the growth-domestic product, whereas the individual network components capture local and global trading patterns that cannot be explained by the covariates alone.
 \end{itemize}
	
    To further motivate our methodology, consider a synthetic network data example in \reffig{fig:sbm:network:1} (the full data generation process is discussed in Section~\ref{Example: data with group structure}) with $n=40$ units (nodes) divided into four groups (indicated by color and shape in \reffig{fig:sbm:network:1}). The nodes are divided into three communities (indicated by shape), with groups 3 and 4 merged into one community. The node covariate data consists of a mixture model with three clusters (indicated by color), where groups 1 and 2 are merged into the same cluster. In this example,  neither the connectivity data nor the covariate data alone can differentiate all four groups. Our methodology allows to discover joint components differentiating the nodes into two larger clusters that are common across two datasets (\reffig{fig:sbm:joint:2}), and individual components that identify the remaining groups that cannot be distinguished by a single view (\reffig{fig:sbm:indiv:3}).

    \begin{figure}[!t] 
		\centering
		\begin{subfigure}[b]{0.4\textwidth}
			\includegraphics[trim={3cm 0cm 0cm 0cm}, clip, scale = .4]{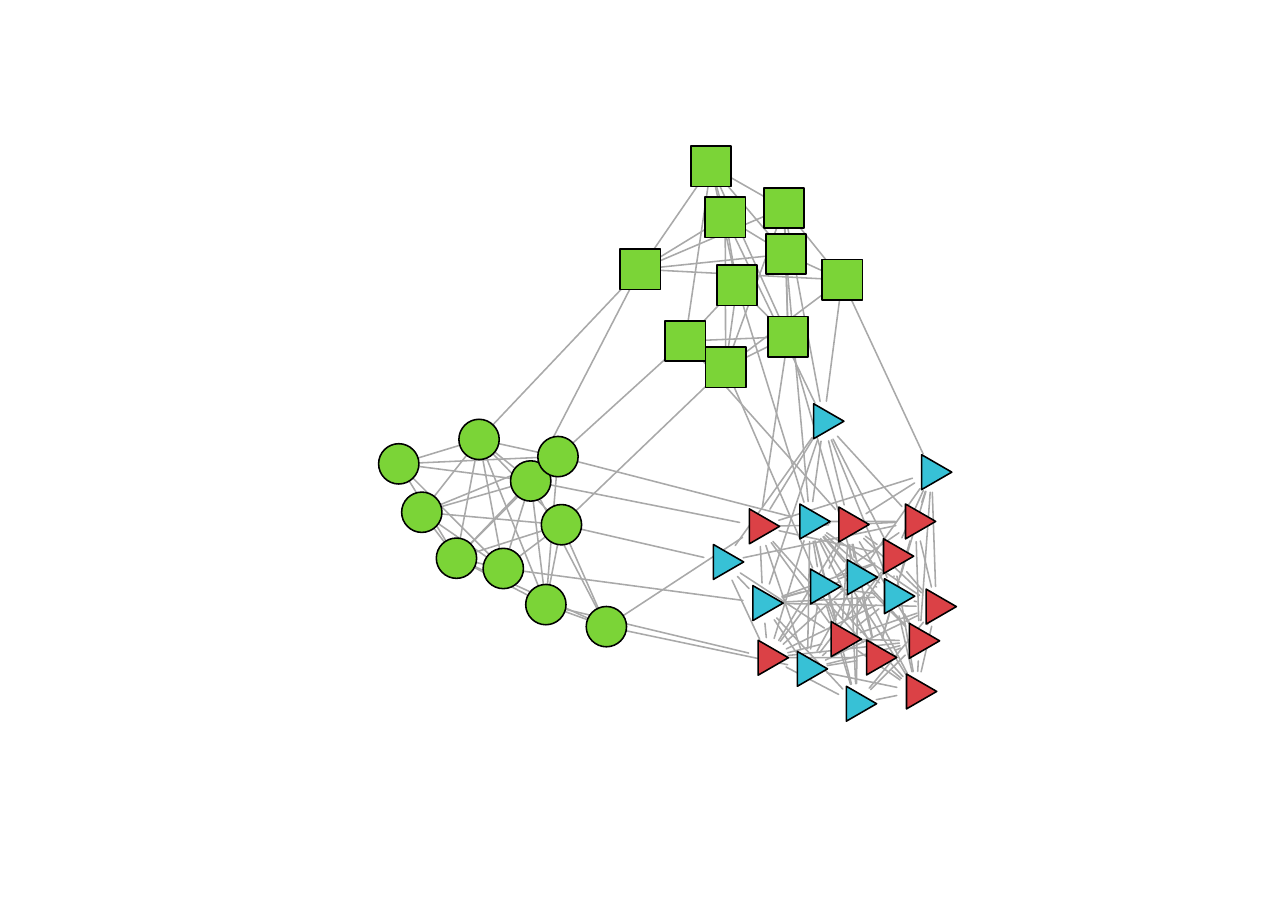}
			\caption{Network connectivity plot}
			\labfig{fig:sbm:network:1}
		\end{subfigure}
		\hfill
		\begin{subfigure}[b]{0.4\textwidth}
			\includegraphics[scale = .4]{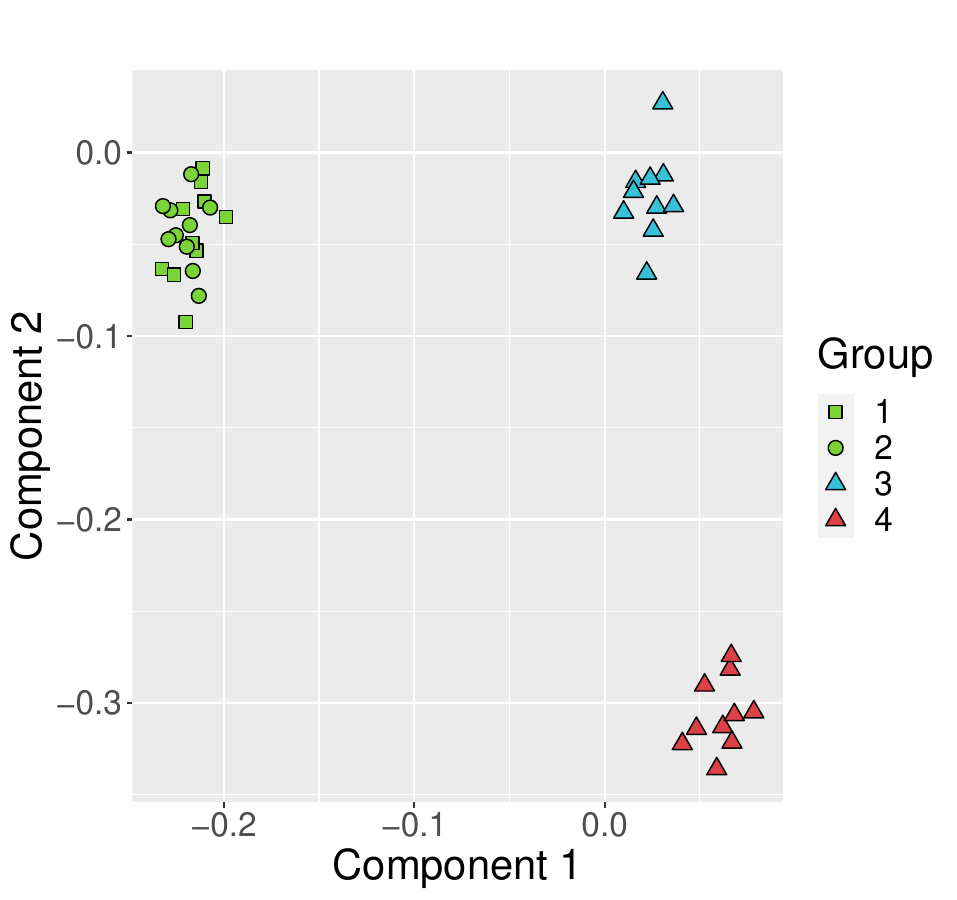}
			\caption{Top node covariate principal components}
			\labfig{fig:sbm:covariates}
		\end{subfigure}
		\hfill
        \begin{subfigure}[b]{0.4\textwidth}
			\includegraphics[scale = .4]{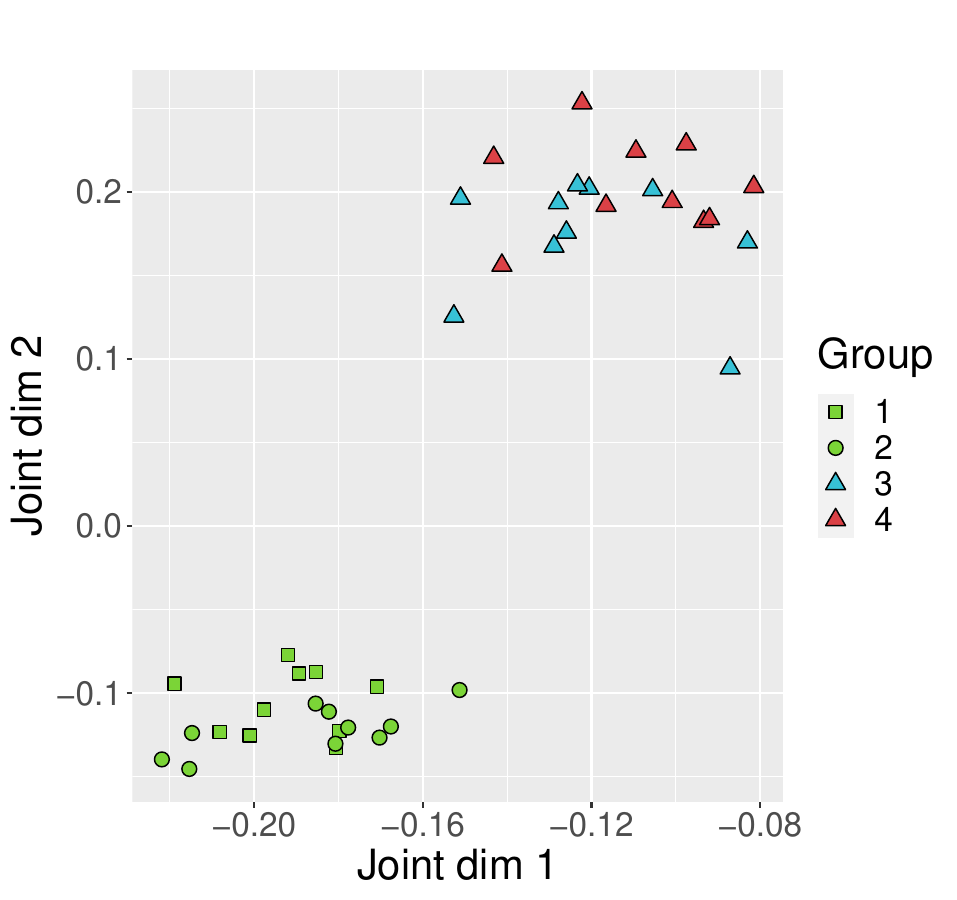}
			\caption{Joint components}
			\labfig{fig:sbm:joint:2}
		\end{subfigure}
        \hfill
        \begin{subfigure}[b]{0.4\textwidth}
			\includegraphics[scale = .4]{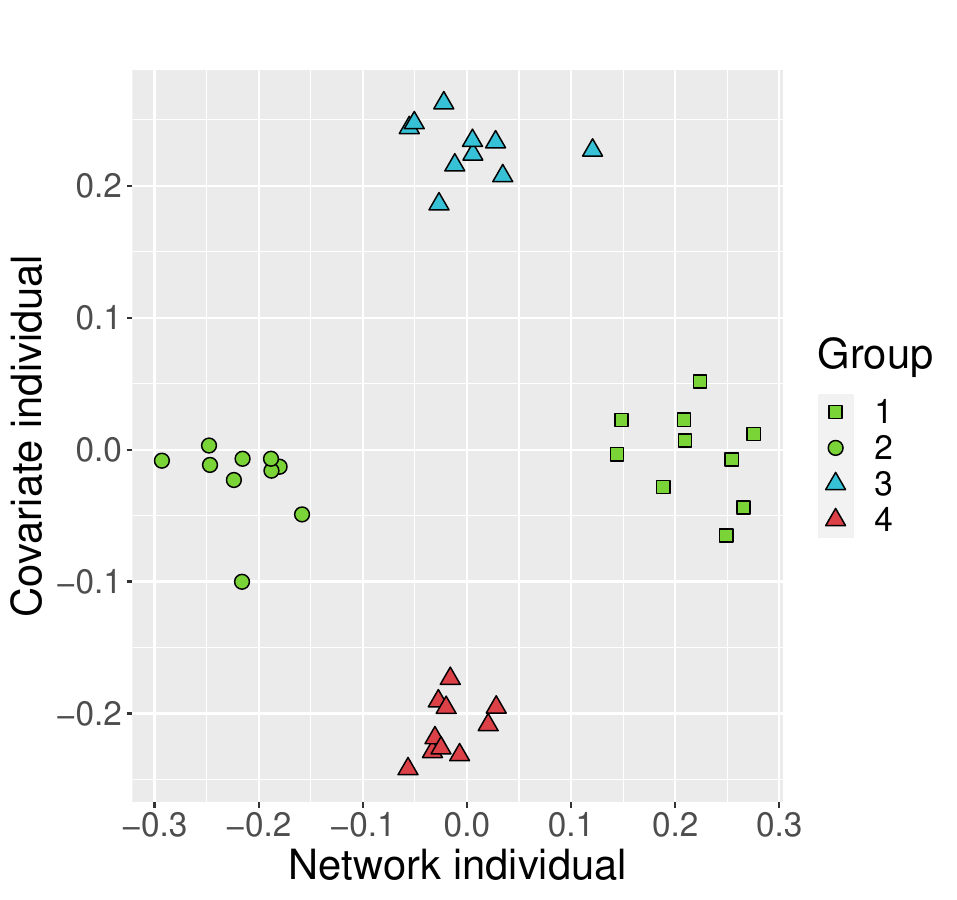}
			\caption{Individual components}
			\labfig{fig:sbm:indiv:3}
		\end{subfigure}
		\caption{Synthetic dataset composed of a network with $40$ nodes (top left) and three node-level covariates represented via their leading principal components (top right). The nodes are partitioned into four groups, but each dataset can only detect three groups.  Shape indicates nodes distinguished by the network data. Color indicates nodes distinguished by covariate data. The joint (bottom left) and individual (bottom right) components estimated by our methodology partition the information into the joint and individual group structures across the datasets. See Section \ref{Example: data with group structure} for details.}
        \labfig{fig:sbm:all}
	\end{figure}

	\subsection{Related Work}
	
	In the analysis of network data, it is common to represent the nodes via a low-dimensional latent space \citep{Hoff:2002:latent_social}. This can be achieved by estimating node embeddings through low-rank decompositions of the adjacency matrix or the Laplacian \citep{Rohe:2011:spec_clust_high_dim_sbm,Athreya:2017:survey}. The resulting low-dimensional representation is often used for downstream inference tasks such as community detection. We follow a similar approach in this work by obtaining low-dimensional representations of the nodes that capture joint and individual variation across network data and covariates.

	In the context of matrix datasets with matched units, there has been substantial work on data integration techniques that appeal to the notion of joint structure. Classical approaches include principal component analysis (PCA) on combined multiple covariate datasets \citep{block_pca_hierarchical, block_pca} and canonical correlation analysis \citep{Harold:1936:cca, Witten:2009:mcca}. More recent approaches decompose the data into joint and individual components \citep{Lock:2013:jive,Feng:2018:ajive,gaynanova2019slide, Murden:2022:ajive_brain,chen2022two, palzer2022sjive}. 
 However, to our knowledge, there persists a gap in the literature for similar integration techniques for network data problems.

	In the context of the analysis of network data with node covariates, much of the existing literature is concerned with two types of approaches. 
    On one hand,  the information in one of the datasets is modeled as a function of the other. For example, the network may be treated as auxiliary information in prediction problems with linked units \citep{Li:2019:predict_network_linked_data,le2022linear}, or alternatively, the edges of the network can be modeled as a function of node and edge covariate information \citep{Hoff:2005:bilinear_mixed_effects}. 
    The second type of approach, which is the direction we follow in this paper, studies both datasets jointly to identify joint structures between the network and its nodal attributes. Common tasks in this approach include community detection with node features \citep{Zhang:2016:comm_dect_node_feat, Binkiewicz:2017:casc,Deshpande:2018:context_sbm,Yan:2021:cov_regularized_community_detection}, or testing for dependence between a network and its nodal attributes\citep{Fosdick:2015:network_covar_dependence,Lee:2019:network_dependence_via_diffusion}. These methods are only concerned with identifying common information across the datasets, but are unable to distinguish between the structure that is shared or unique to each dataset, which is our main goal in our analysis. Preparing this work, we became aware of recent concurrent work that also introduced a notion of joint and individual in network data with covariates \citep{Li:2024:network_linked_data}, but the model and methodology are substantially different from ours.

    Another related line of work is the study of multiple network datasets with matched nodes, often referred to as multilayer or multiplex networks. A common approach in this setup is to identify joint structure across the networks via node embeddings \citep{Eynard:2015:multi_mani_simultaneous_diag, Wang:2019:joint_embedding_graphs, zhang2018relationships, spectral_factorization, Wang:2019:common_indiv_structure, Arroyo:2021:cosie}, but these methods are typically unable to identify individual structures. Recent progress in this direction has been made by \citet{MacDonald:2022:multiplex_latent_space} who partition the model structure into joint and individual parts, but extensions to network data with covariates are not immediate. Although our model shares some similarities with \cite{MacDonald:2022:multiplex_latent_space}, the approach we follow considers modeling the column spaces of the matrices instead of directly modeling the latent factors as in their work, resulting in more flexibility to handle differences in scaling or rotation factors across the datasets. 
    
    From an algorithmic perspective, our methodology is based on a combination of an efficient spectral algorithm to jointly embed the network and the covariates, followed by an iterative optimization procedure as a refinement of the initial estimates. Our spectral method shares some similarities with the AJIVE method of \citet{Feng:2018:ajive} for matrix data, but our model is fundamentally different as it treats network data with a connection to a latent space model. Furthermore, we provide theoretical guarantees on the error rate of the spectral method, which are still not available in the matrix setup. The spectral method is also related to algorithms for finding common low-rank embeddings \citep{Crainiceanu:2011:population_value_decomp,Arroyo:2021:cosie}, but the latter methods are unable to recover individual components.

	\subsection{Notation.}
	The column space of a matrix $M \in \R^{n \times m}$ is denoted by $\MC(M)$, and the orthogonal complement of the subspace $\MC(M)$ is denoted by $\MC(M)^{\perp}$. The set of $n\times p$ matrices with orthonormal columns is denoted by $\O_{n, p} = \{Q \in \R^{n \times p}: Q^{\top}Q = I\}$, and we write $\O_n$ in place of $\O_{n,n}$. Given matrices $U, V \in \R^{n \times m}$, we denote by $U \cong_{O} V$ if there exists $W\in\O_{p}$  such that $U = VW$. Given a symmetric matrix $A \in \R^{n \times n}$ with eigendecomposition $U \Lam U^{\top}=A$, where $U = (u_1, \ldots, u_n) \in \O_{n}$ contains the eigenvectors and $\Lam = \text{diag}(\lam_1, \ldots, \lam_n)$ contains the eigenvalues in its diagonal with $|\lam_1| \geq \cdots \geq |\lam_n|$, we denote the matrix of $k$ leading eigenvectors of $A$ ordered by eigenvalue magnitude by $\eig(A, k) = (u_1, \ldots, u_k) \in \R^{n \times k}$.
	Given a matrix $X \in \R^{n \times p}$, let $X=U\Sigma V^\top$ be its singular value decomposition, where $U = (u_1, \ldots, u_m) \in \O_{n\times m}$, $V\in\O_{p\times m}$, and $\Sigma = \text{diag}(\sig_1, \ldots, \sig_m)$ with $\sig_1 \geq \cdots \geq \sig_m>0$. 
	The $j$-th largest singular value of $X$ is denoted by $\sig_j(X)$, and the $k$ left leading singular vectors of $X$ ordered by singular value magnitude are denoted by $\sv(X, k) = (u_1, \ldots, u_k) \in \R^{n \times k}$.  Let $f,g:\N \rightarrow [0, \infty)$. We write $f = O(g)$ and $g = \Om(f)$ if there exists $M>0$ and $n_0 \in \N$ such that for each $n \in \N$, $n \geq n_0$ implies that $f(n) \leq M g(n)$.  We write $f \asymp g$ if $f = O(g)$ and $f= \Om(g)$. We write $f = o(g)$ and $g = \om(f)$ if for each $\ep >0$, there exist $n_0 \in \N$ such that for each $n \in \N$, $n \geq n_0$ implies that $f(n) < \ep g(n)$. We denote the Frobenius and spectral norms by $\| \cdot \|_F$ and $\| \cdot \|$ respectively.

	
	\section{Methodology}
        In this section, we propose a model for network data with covariates, and develop estimation procedures for the model parameters.
	\subsection{Modeling joint and individual structure} \label{Model}

 We consider the problem of jointly modeling the structure of a network (a graph with $n$ nodes) and node-level covariate information. We represent the network by an observed adjacency matrix $A \in \R^{n \times n}$, where the entry $A_{ij}$ encodes the value of the interaction between nodes $i$ and $j$. We represent node-level covariates by a matrix $X \in \R^{n \times p}$ so that the $i$-th row of $X$ represents $p$ variables measured at node $i$. For simplicity, we focus on the undirected network setup, in which case $A$ is symmetric. 

We adopt a signal-plus-noise matrix model for the adjacency matrix $A\in\R^{n\times n}$ and the covariate matrix $X\in\R^{n\times p}$ so that
\begin{align} 
    & A = P + E^A, \quad \quad  X = W + E^X,\labeq{model:sig+noise:1}
\end{align} 
where $P=\E[A] \in \R^{n \times n}$ and $W=\E[X] \in \R^{n \times p}$ are the expected values, $E^A$ and $E^X$ are random matrices such that the collection of upper triangular entries of $E^A$ and the entries of $E^P$ are mutually independent random variables, and $P$ and $E^A$ are symmetric matrices. The matrix $P$ represents the expected node connection strengths (network signal), whereas $W$ represents the signal of the node covariates. The additive decomposition for $A$ incorporates many existing models. For instance, when elements of $A$ are binary, it is common to model its entries as independent Bernoulli random variables with $A_{ij} {\sim} \text{Bern}(P_{ij})$, where $P_{ij} \in [0,1]$ indicates the probability of connection between nodes $i$ and $j$.

   To model joint and individual structures across the network and covariate signals, we propose to partition the column spaces of signal matrices $P$ and $W$ such that the joint structure is the intersection of these spaces, and the individual structures are the residuals after removing the intersection \citep{Lock:2013:jive,Feng:2018:ajive}.  We formalize this with the definition below.

    \begin{defn} \labdef{def:model:latent_positions:1}
		Let $P \in \R^{n \times n}$ and $W \in \R^{n \times p}$ be the expected values of the network adjacency matrix and node covariate matrix from \refeq{model:sig+noise:1}. We define the \emph{joint subspace}, \emph{individual subspace for the network} and the \emph{individual subspace for the covariates} as
    \begin{align} \labeq{model:joint_individual_subspaces:2.01}
			& \text{\emph{Joint:}} \quad  &\MM  &= \MC(P) \cap \MC(W), \\
	 \labeq{model:joint_individual_subspaces:3.01}
			& \text{\emph{Network Individual:}} \quad  &\MR^{(1)} &= \MC(\MP_{\MM^{\perp}}P), \\
    \labeq{model:joint_individual_subspaces:4.01}
			 & \text{\emph{Covariate Individual:}}  \quad  &\MR^{(2)} &= \MC(\MP_{\MM^{\perp}}W). 
	\end{align}
		 We denote the respective dimensions of these subspaces by $r_M$, $r_1$ and $r_2$. If $M \in \O_{n, r_M}$, $R^{(1)} \in \O_{n, r_1}$ and $R^{(2)} \in \O_{n, r_2}$ satisfy $\MC(M) = \MM$, $\MC(R^{(1)}) = \MR^{(1)}$ and $\MC(R^{(2)}) = \MR^{(2)}$, we refer to $M$, $R^{(1)}$ and $R^{(2)}$ as the \emph{joint components}, \emph{network individual components} and \emph{covariate individual components} respectively.
	\end{defn}

    The joint and individual subspaces in \refdef{def:model:latent_positions:1} are uniquely defined by construction and satisfy $\rnk(P) = r_M + r_1$ and $\rnk(W) = r_M + r_2$. While the subspaces are unique, the joint and individual components $M, R^{(1)}, R^{(2)}$ are only unique up to multiplication by orthogonal matrix. We formalize the uniqueness conditions as well as decomposition of $P$ and $W$ through these components below under an additional nontriviality assumption.

    \begin{asmp}  
    \label{model:nontriviality:number}
    \labasmp{model:nontriviality:1}
    The joint and individual subspaces $\MM, \MR^{(1)}$ and $\MR^{(2)}$ from \refdef{def:model:latent_positions:1} are non-trivial, that is, $\MM, \MR^{(1)},\MR^{(2)} \neq \{0\}$. 
    \end{asmp}
    \noindent
\refasmp{model:nontriviality:1} implies that $r_M$, $r_1$, $r_2 > 0$, meaning that each subspace in Definition~\ref{defn:def:model:latent_positions:1} is non-empty.

	\begin{lem} \lablem{lemma:model:identifiability:1} 
		Let $P \in \R^{n \times n}$ and $W \in \R^{n \times p}$ be as in \refeq{model:sig+noise:1} and let $r_M, r_1, r_2$ be the ranks from  \refdef{def:model:latent_positions:1}. Under \refasmp{model:nontriviality:1}, there exist $M \in \O_{n , r_M}$, $R^{(k)} \in \O_{n , r_k}$ for $k = 1,2$, as in \refdef{def:model:latent_positions:1}, and full rank matrices $\Gam^{(1)} \in \R^{(r_M + r_1) \times n}$ and $\Gam^{(2)} \in \R^{(r_M + r_2) \times p}$ such that
		\begin{enumerate}[(a)]
            \item $P = 
			\begin{pmatrix}
				M & R^{(1)}
			\end{pmatrix} 
			\Gam^{(1)}$ and
            $W = 
			\begin{pmatrix}
				M & R^{(2)}
			\end{pmatrix} 
			\Gam^{(2)}$,
             \item $M \perp R^{(k)}$, $k=1,2$,
			\item $M,R^{(1)}$ and $R^{(2)}$ are unique up to right multiplication by an orthogonal matrix.
		\end{enumerate}
	\end{lem}
 
	Lemma~\ref{lemma:lemma:model:identifiability:1} establishes the identifiability of the components up to an orthogonal matrix multiplication, and shows that the joint and individual components are orthogonal to each other.  Related results for matrix-valued data have been studied in \citet{Feng:2018:ajive, Yuan:2022:DMMD}. The decomposition in (a) characterizes the column spaces of the matrices in terms of such components. As a consequence,  the network  components 
$\begin{pmatrix}
M & R^{(1)}
\end{pmatrix}$ are equivalent to the non-trivial eigenvectors of $P$ up to an orthogonal transformation, whereas the covariate components
$\begin{pmatrix}
M & R^{(2)}
\end{pmatrix}$
are the left singular vectors of $W$ up to an orthogonal transformation, which motivates us to utilize a spectral method for estimation (see Section~\ref{spec_est_section}).

 \refdef{def:model:latent_positions:1} relates to several existing methods. When the network adjacency matrix is replaced by a non-symmetric matrix of covariates (no network data), \refdef{def:model:latent_positions:1} coincides with existing models for joint and individual signals for matrix-valued data \citep{Lock:2013:jive,Feng:2018:ajive}. In comparison to existing approaches for network data, the special case of $r_1=r_2 = 0$  (column spaces of $P$ and $W$ are the same) is similar to models that consider common latent factors or community structure \citep{Binkiewicz:2017:casc, Arroyo:2021:cosie}; thus, our model is more general. A recent work by \cite{MacDonald:2022:multiplex_latent_space} considers a similar decomposition in terms of both joint and individual components for multiplex networks. However, their model imposes this decomposition in the factorization of the latent positions directly. In contrast, we use the column spaces, which allow for more flexibility via different scaling and rotation factors across the components of  $P$ and $W$.

For estimation to be feasible, we assume that the signal matrices $P$ and $W$ are low rank, that is $n \gg r_M + r_1$, $\min(n, p) \gg r_M +  r_2$, which is a common assumption for network and matrix data \citep{Athreya:2017:survey,Han:2023:univ_rank_infer,Udell:2019:Why_are_big_mat_low_rank}. However, for network data, the low-rank assumption may not hold exactly but only approximately. For a simple example, if the network has no self-loops, the diagonal entries of $P$ are zero, which can violate the low-rank assumption.
     In these cases, \refdef{def:model:latent_positions:1} can be modified to work with the leading eigenspace of $P$ and the leading
    singular subspace of $W$. Formally, if $V^{(1)}\in\O_{n, r_P}$ is the matrix of leading eigenvectors of $P$ ($V^{(1)} = \eig(P, r_P)$) and $V^{(2)}\in\O_{n, r_W}$ is the matrix of left-leading singular vectors of $W$ ($V^{(2)}=\sv(W, r_W)$) for any given values of $r_P$ and $r_W$, the joint and individual subspaces can be defined in an analogous manner as $\widetilde{\MM} = \MC(V^{(1)})\cap  \MC(V^{(2)})$, $\widetilde{\MR}^{(1)} = \MC(\MP_{{\widetilde{\MM}}^{\perp}}V^{(1)})$ and $\widetilde{\MR}^{(2)} = \MC(\MP_{{\widetilde{\MM}}^{\perp}}V^{(2)})$, respectively. This definition coincides with \refdef{def:model:latent_positions:1} when $r_P = \rnk(P)$ and $r_W = \rnk(W)$. We opted to work with a low-rank assumption for simplicity of notation, but all our results continue to hold if the matrices are approximately low-rank.

\subsubsection{Implications for latent space models with covariates} \label{Latent Space}

While our model in  \reflem{lemma:model:identifiability:1} is stated in terms of the column space of the network, it has a connection to latent space models \citep{Hoff:2002:latent_social} and implies a decomposition of the latent positions of the network into joint and individual parts. Concretely, consider the \emph{generalized random dot product graph} (GRDPG) model \citep{Rubin-Delanchy:2022:GRDPG}. This model represents the nodes of the network with latent positions in some low-dimensional space and characterizes the probability of two nodes being connected by an edge as an indefinite inner product between their corresponding latent positions. This model is particularly suitable to characterize networks with low-rank expected adjacency matrices, and encompasses multiple models of interest in the literature, including the stochastic blockmodel \citep{Holland:1983:stochastic_block_model} and its extensions. For a survey, see \cite{Athreya:2017:survey}.

Specifically, for a binary and symmetric adjacency matrix $A$ whose edges $(A_{ij})_{i < j}$ are independent, the GRDPG model assigns to each node $i=1, \ldots, n,$
 a corresponding latent position denoted by $y_i \in \R^{d}$, and the probability of a connection between nodes $i$ and $j$ is given by the indefinite inner product of their associated latent positions $P_{ij}  = y_i^{\top} I_{a,b} y_j$,  where $I_{a,b} = \text{diag}(1, \ldots, 1, -1, \ldots, -1)$ is a $d\times d$ diagonal matrix with $a$ ones and $b$ minus ones on its diagonal such that $a+b = d$. We collect these latent positions into a matrix 
	$Y = 
	\begin{pmatrix}
		y_1, \cdots, y_n 
	\end{pmatrix}^{\top} \in \R^{n \times d}$ so that the expected adjacency matrix can be expressed as $P = Y I_{a,b}Y^{\top}$. We denote $A \sim \operatorname{GRDPG}(Y, a, b)$ for a network adjacency matrix generated under this model. If $b = 0$, this model reduces to the \emph{random dot product graph} (RDPG) \citep{Young:2007:RDPG} and we write $A \sim \operatorname{RDPG}(Y)$.

When a graph is modeled jointly with node covariates as in \refdef{def:model:latent_positions:1}, the latent positions of the GRDPG model can be decomposed into joint and individual components. In particular, the column space of $P$ is the same as that of the latent position matrix, that is, $\MC(P) = \MC(Y)$. Using the factorization for $P$ given in \reflem{lemma:model:identifiability:1}, we can decompose the latent positions $Y$ into joint and individual parts as demonstrated next.

\begin{cor}
Let $P$ and $W$ be matrices as in \refeq{model:sig+noise:1} such that $P$ is the probability matrix of a generalized random dot product graph model $\operatorname{GRDPG}(Y, a, b)$ with latent position matrix $Y \in \R^{n\times d}$. Suppose that $M, R^{(1)}$ and $R^{(2)}$ are the joint and individual components from \refdef{def:model:latent_positions:1}. Then, there exists a matrix $\widetilde{\Gamma}^{(1)} \in \R^{(r_M+r_1) \times d}$ such that
\begin{equation}
\labeq{eq:model:latent_pos}
Y = \begin{pmatrix}
				M & R^{(1)}
			\end{pmatrix} \widetilde{\Gamma}^{(1)} = \MP_{\MC(M)} Y + \MP_{\MC(R^{(1)})} Y.
\end{equation}
 \end{cor}
 This corollary is a direct consequence of Lemma~\ref{lemma:lemma:model:identifiability:1}. By virtue of the factorization in \refeq{eq:model:latent_pos},  we can interpret the rows of $M$ as the node embeddings that explain the joint structure with the covariate information, whereas the rows of $R^{(1)}$ correspond to latent positions individual to the network. For example, in the stochastic blockmodel \citep{Holland:1983:stochastic_block_model} (SBM), which is a special case of the GRDPG model, the rows of $Y$ take only $K$ different values, corresponding to the clusters or communities of the model, and the decomposition in \refeq{eq:model:latent_pos} breaks these $K$ groups according to the structure in the covariates. In particular, if the covariate data also presents clustering structure (as illustrated in \reffig{fig:sbm:network:1}), the joint and individual components can detect clustering structure that is common or individual across the two datasets. We elaborate an example of this situation in Section~\ref{Example: data with group structure}.

	\subsection{Estimation} \label{estimation_section}
	In this section, we describe our methodology for estimating joint and individual components in network data with covariates as defined in \refdef{def:model:latent_positions:1}. We first obtain estimates via a simple and efficient spectral method. We then refine these initial estimates via joint optimization to obtain the final estimates. Throughout, we assume that the ranks $r_M$, $r_1$ and $r_2$ are known. In our application, we estimate the ranks by finding elbows in scree plots (Section ~\ref{data_analysis}), but alternative methods for rank estimation based on singular value thresholding can be used \citep{Chatterjee:2015:USVT, Gavish:2017:opti_shrink_sing_vals, prothero:2024:DIVAS}. 
	
	\subsubsection{Spectral Method} \label{spec_est_section}
	To motivate the spectral method, we start by focusing on the noiseless case, in which the expectation matrices $P$ and $W$ are directly observed. We set $V^{(1)} = \sv(P, r_M + r_1)$ and $V^{(2)} = \sv(W, r_M + r_2)$ as the leading eigenvector and left singular vector matrices of $P$ and $W$, respectively. Since $V^{(k)}$ and $
    \begin{pmatrix}
        M & R^{(k)}
    \end{pmatrix}$ have the same column space, $V^{(k)} \cong_O 
    \begin{pmatrix}
        M & R^{(k)}
    \end{pmatrix}$. 
	The key idea behind the spectral estimation procedure is to combine the datasets in such a way that the joint subspace has a stronger signal in the combination than in either of the original datasets. To this end, we consider the matrix $U = 
	\begin{pmatrix}
		V^{(1)} & V^{(2)} 
	\end{pmatrix}$ and observe that its $r_M$ leading left singular values correspond to $M$ (modulo some orthogonal matrix). Once we have $M$, we can obtain $R^{(k)}$ using the fact that 
    $$\sv(\MP_{\MC(M)^{\perp}} V^{(k)}, r_k) \cong_O \sv(\begin{pmatrix}
        0 & R^{(k)}
    \end{pmatrix}, r_k) \cong_O R^{(k)}.$$ 
    In practice, as the matrices $P$ and $W$ are not observed, we apply this process to observed $A$ and $X$ as detailed in \refalg{alg:model:spectral:1}. When applied to the noiseless $P$ and $W$, the algorithm indeed recovers the true model parameters $M, R^{(1)}, R^{(2)}$, as formalized below.
    
	\begin{prop}\labprop{proposition:estimation:spectral:1}
		 Let $P \in \R^{n \times n}$ and $W \in \R^{n \times p}$ be the signal matrices as in \refeq{model:sig+noise:1}, and let $r_M, r_1, r_2$ be the ranks as in  \refdef{def:model:latent_positions:1}. Suppose that \refalg{alg:model:spectral:1} is started with $(P, W, r_M, r_1, r_2)$ as the input. Then, the algorithm returns the matrices $M, R^{(1)}, R^{(2)}$ that satisfy Lemma~\ref{lemma:lemma:model:identifiability:1}, up to right multiplication by an orthogonal matrix.
	\end{prop}
    
    \begin{algorithm}[!t] 
		\caption{Spectral method for joint and individual component estimation} 
		\labalg{alg:model:spectral:1}
		\begin{algorithmic} 
			\Input $A \in \R^{n \times n}$, $X \in \R^{n \times p}$, $r_M$, $r_1$, $r_2$ \vspace{.1cm}
			\State \quad \quad $\widehat{V}^{(1)} = \eig(A, r_M + r_1)$ \vspace{.1cm}
			\State \quad \quad $\widehat{V}^{(2)} = \sv(X, r_M + r_2)$ \vspace{.1cm}
			\State \quad \quad $\widehat{U} = 
			\begin{pmatrix}
				\widehat{V}^{(1)} & \widehat{V}^{(2)}
			\end{pmatrix}$ \vspace{.1cm}
			\State \quad \quad $\widehat{M} = \sv(\widehat{U}, r_M)$ \vspace{.1cm}
			\State \quad \quad  $\widehat{R}^{(1)} = \sv(\MP_{\MC(\widehat{M})^{\perp}} \widehat{V}^{(1)}, r_1)$ \vspace{.1cm}
			\State \quad \quad  $\widehat{R}^{(2)} = \sv(\MP_{\MC(\widehat{M})^{\perp}} \widehat{V}^{(2)}, r_2)$ \vspace{.1cm}
			\Output {$\widehat{M}, \widehat{R}^{(1)}, \widehat{R}^{(2)}.$} \vspace{.1cm}
		\end{algorithmic}
	\end{algorithm}
    
	\subsubsection{Optimization Algorithm Refinement} \label{optimization_refinement}
	
	While the spectral method in Section~\ref{spec_est_section} is simple to implement and computationally fast, it can discard nontrivial signals in the initial low-rank approximation step. To remedy this, we propose a refinement estimation procedure based on optimizing an objective function associated with the model. To measure the loss for the covariates, we consider the standard squared Frobenius norm error, $\|X - W\|_F^2$. For the network, due to the symmetry of the adjacency matrix, we consider non-symmetric matrices $A' = (|\widehat{\lam}_1|^{1/2} \widehat{u}_1, \ldots, |\widehat{\lam}_n|^{1/2} \widehat{u}_n)$ and $P' = (|\lam_1|^{1/2} u_1, \ldots, |\lam_n|^{1/2} u_n)$ where $\widehat{\lam}_1, \ldots, \widehat{\lam}_n$ and $\widehat{u}_1, \ldots, \widehat{u}_j$ are the eigenvalues and eigenvectors of $A$ ordered by magnitude, and $\lam_1, \ldots, \lam_n$ and $u_1, \ldots, u_j$ are the eigenvalues and eigenvectors of $P$ ordered by magnitude. Since $A$ and $A'$ have the same column space (similarly, $P$ and $P'$) the conditions as in \reflem{lemma:model:identifiability:1} still apply. When $A$ is positive-definite, $A'$ and $P'$ are the square roots of $A$ and $P$, respectively. We consider the Frobenius norm error $\|A' - P'\|_F^2$ to measure the loss for the network.

To further refine the initial estimates obtained from \refalg{alg:model:spectral:1}, we propose to minimize the sum of the losses associated with the network and covariate signals subject to identifiability conditions in \reflem{lemma:model:identifiability:1}. Specifically, since $\MC(P')=\MC(P)=\MC(M\  R^{(1)})$ and $\MC(W)=\MC(M\ R^{(2)})$, under the identifiability conditions the signals minimizing Frobenius norm error loss must satisfy
$$\wh{P'} = \MP_{\MC(\wh{M})}A' + \MP_{\MC(\wh{R}^{(1)})} A', \quad \quad \wh{W} = \MP_{\MC(\wh{M})}X + \MP_{\MC(\wh{R}^{(2)})}X.
$$
This leads us to the following constrained optimization problem
	\begin{align}
		\label{optim:model:main_optim:1}
		\minimize_{M, R^{(1)}, R^{(2)}}& \left  \{\| A' - \MP_{\MC(M)}A' - \MP_{\MC(R^{(1)})} A'\|_F^2 + \|X - \MP_{\MC(M)}X - \MP_{\MC(R^{(2)})}X\|_F^2 \right \} \\
		\mbox{s.t.} & \quad \begin{pmatrix}
				M & R^{(k)}
			\end{pmatrix} \in \O_{n , r_M + r_k}.
	\end{align}
The optimization problem above seeks the best low-rank approximations to the observed data matrices simultaneously under the joint subspace constraint imposed by the model.  To ensure equal weights of both terms in the loss function, we scale $A'$ and $X$ by dividing all elements by the Frobenius norm of $\sv(A', r_M + r_1)$ and $\sv(X, r_M + r_2)$, respectively. In the remainder of this section, we assume that $A'$ and $X$ have already been scaled.

To solve the optimization problem \eqref{optim:model:main_optim:1}, we use the initial estimates obtained from~\refalg{alg:model:spectral:1}, and perform iterative updates. That is, we fix $M$ and optimize for $R^{(k)}$, $k=1,2$, and subsequently fix $R^{(k)}$ and optimize for $M$. These updates are alternated until convergence. We describe each update below.

	When $M$ is fixed, by Lemma~2 in \citet{Yuan:2022:DMMD}, the updates for $R^{(1)}$ and $R^{(2)}$ are closed form and are given by
    $$\wh{R}^{(1)} = \sv( \MP_{\MC(M)^{\perp}} A', r_1), \quad \quad \wh{R}^{(2)} = \sv( \MP_{\MC(M)^{\perp}} X, r_2).$$
	When $R^{(1)}$ and $R^{(2)}$ are fixed, we derive the closed-form update for $M$ based on the following lemma.

 \begin{lem}
\lablem{lemma:loss_function_minimizer:1}
Consider optimization problem~\eqref{optim:model:main_optim:1} with respect to $M$ with fixed $R^{(k)} \in \O_{n, r_k}$ for $k=1, 2$ and fixed joint rank $r_M$. Let $Y = 
		\begin{pmatrix}
			\MP_{\MC(R^{(1)})^{\perp}} A' & \MP_{\MC(R^{(2)})^{\perp}} X
		\end{pmatrix}$ and $R = 
		\begin{pmatrix}
			R^{(1)} & R^{(2)}
		\end{pmatrix}$.
		 Set $\wh{M} = \sv(\MP_{\MC(R)^{\perp}} Y, r_M)$. If $\rnk (\MP_{\MC(R)^{\perp}} Y )\geq r_M$, then $\wh{M}$ is a global minimizer of~\eqref{optim:model:main_optim:1}.
 \end{lem}
	
	In \reflem{lemma:loss_function_minimizer:1}, we assume $\MP_{\MC(R)^{\perp}} Y$ is of rank at least $r_M$. If $R=
		\begin{pmatrix}
			R^{(1)} & R^{(2)}
		\end{pmatrix}$ 
		contains a complete basis for $\R^n$, then $\MP_{\MC(R)^{\perp}} Y = 0$, which means there is no solution for $M$ with given rank $r_M$. In practice, we never encounter such situations as the signal ranks are smaller than the maximal ranks, and our starting values for $M$, $R^{(1)}$ and $R^{(2)}$ always satisfy the assumption.
  
  \refalg{alg:model:optimization:2} summarizes overall iterative updates for problem~\eqref{optim:model:main_optim:1}.
  While each block update has a closed-form solution, the final output can be affected by the initial starting points as the overall problem is nonconvex. We purposefully use our initial estimates obtained from \refalg{alg:model:spectral:1} as a starting point, and thus Algorithm~\eqref{optim:model:main_optim:1} can be viewed as an optimization-based refinement of the spectral estimates. In Section~\ref{Estimation Error of Spectral Method}, we further prove that the initial estimates are consistent.

	\begin{algorithm}[!t]
		\caption{Iterative descent method for joint and individual component estimation}
		\labalg{alg:model:optimization:2} 
		\begin{algorithmic} 
			\Input Scaled data matrices, $A' \in \R^{n\times n}$, $X \in \R^{n\times p}$; initial estimate $M_{(0)} \in \R^{n\times r_M}$; $r_1, r_2, t_{max},\epsilon > 0$. 
			\Initialize 
            \State \quad $t = 0$;  $L_0 = 0$ \vspace{.1cm}
			\State \quad $R^{(1)}_{(0)} = \sv (\MP_{\MC(M_{(0)})^{\perp}}A', r_1)$ \vspace{.1cm}
			\State \quad $R^{(2)}_{(0)} = \sv(\MP_{\MC(M_{(0)})^{\perp}} X, r_2)$ \vspace{.1cm} 
			\Repeat 
			\State Update $M$: \vspace{.1cm}
			\State \quad \quad $Y_{(t)} = 
			\begin{pmatrix} 
				\MP_{\MC(R^{(1)}_{(t)})^{\perp}} A' & \MP_{\MC(R^{(2)}_{(t)})^{\perp}} X
			\end{pmatrix}$ \vspace{.1cm}
			\State \quad \quad  $R_{(t)} = 
			\begin{pmatrix}
				R^{(1)}_{(t)} & R^{(2)}_{(t)}
			\end{pmatrix}$ \vspace{.1cm}
			\State \quad \quad $M_{(t+1)} = \sv( \MP_{\MC(R_{(t)})^{\perp}} Y_{(t)}, r_M)$ \vspace{.1cm}
			\State Update $\wt{R}^{(1)}$ and $\wt{R}^{(2)}$: \vspace{.1cm}
			\State \quad \quad $R^{(1)}_{(t+1)} = \sv( \MP_{\MC(M_{(t+1)})^{\perp}} A', r_1)$ \vspace{.1cm}
			\State \quad \quad $R^{(2)}_{(t+1)} = \sv( \MP_{\MC(M_{(t+1)})^{\perp}} X, r_2)$ \vspace{.1cm}
			\State Update iteration and loss: \vspace{.1cm}
			\State \quad \quad $t = t+1$  \vspace{.1cm}
			\State \quad \quad  $V^{(k)}_{(t)} = 
			\begin{pmatrix}
				M_{(t)} & R^{(k)}_{(t)}
			\end{pmatrix}$ \hspace{.2cm } $k = 1,2$ \vspace{.1cm}
			\State \quad \quad $L_{(t)} = \|A' -  \MP_{V^{(1)}_{(t)}} A'\|_F^2 + \|X - \MP_{V^{(2)}_{(t)}} X\|_F^2$ \vspace{.1cm}
			\Until{{$t > t_{max}$ or $|L_{(t)} - L_{(t-1)}| \leq \epsilon$} \vspace{.1cm} }
			\State \quad $\wh{M} = M_{(t)}$ \vspace{.1cm}
			\State \quad $\wh{R}^{(k)} = R^{(k)}_{(t)}$ \hspace{.2cm } $k = 1,2$ \vspace{.1cm}
			\Output {$\wh{M}$, $\wh{R}^{(1)}$, $\wh{R}^{(2)}$}
		\end{algorithmic}
	\end{algorithm}

	\section{Estimation Error of Spectral Method} \label{Estimation Error of Spectral Method}
	
	In this section, we characterize the estimation error of the spectral method from Section~\ref{spec_est_section}. Throughout the analysis, we assume the true ranks $r_M, r_1, r_2$ are known. To quantify the estimation error, we use the Procrustes distance which is invariant to orthogonal transformations (\reflem{lemma:model:identifiability:1}). For matrices with orthonormal columns $U,V \in \O_{n , r}$, it is defined as
	\begin{equation}\labeq{eq:procrustes_distance}
        d(U, V) = \inf_{Q \in \O_r} \|U - VQ\|_F.
    \end{equation} 
    Our goal is to characterize the expected distance between the true and estimated components, that is $\E[ d(\wh{M}, M)]$ and $\E[ d(\wh{R^{(k)}}, R^{(k)})]$, $k=1, 2$. We will obtain such characterization by leveraging perturbation theory for eigenspaces and singular subspaces \citep{Yu:2015:davis_kahan_for_stats, Cai:2018:rate_optim_perturb_bound_sing_subspace} under the following distributional assumptions. 
	
    \begin{asmp} 
    \label{asmp:estimation:distribution_general:number}\labasmp{asmp:estimation:distribution_general}
    Let $A \in \R^{n \times n}$ and $X \in \R^{n \times p}$ be the observed adjacency and covariate matrices with expectation matrices $P$ and $W$ as in \refeq{model:sig+noise:1} 
    The following distributional assumptions hold. 
    \begin{enumerate}[label=(\alph*)]
    \item $\|A - P\|$ is a subexponential random variable with parameter $\kappa > 0$, that is, for each $q \geq 1$, $\E(\|A-P\|^q)^{1/q} \leq \kappa q$.
    
    \item The entries of $X$ are independent subgaussian random variables with $\text{Var}(X_{ij}) = \tau^2$, that is, there exists $c > 0$ such that for each $t \geq 0$, $\mathbb{P}(\|X_{ij}\| \geq t) \leq 2 e^{-t^2/c}$.
    
    \end{enumerate}
    \end{asmp}
    \noindent
    Many existing network models satisfy this assumption, including the GRDPG with binary entries discussed in Section \ref{Latent Space}.

	One of the main difficulties in estimating the components from \refdef{def:model:latent_positions:1} is distinguishing the joint components from individuals, as the latter are not necessarily orthogonal. When the individual subspaces are closely aligned, they can be mistaken for joint subspaces in the presence of noise, making estimation difficult. To provide a theoretical characterization of this difficulty, we use the first principal angle  between the individual subspaces $\MR^{(1)}$ and $\MR^{(2)}$ to quantify their separation as
    \begin{equation}
   \delta = \del(\MR^{(1)}, \MR^{(2)}) = 1- \cos\{\theta_1(\MR^{(1)}, \MR^{(2)})\} = 1 - \sig_1({R^{(1)}}^{\top} R^{(2)}),
    \labeq{eq:estimation:delta}
    \end{equation}
    where $R^{(1)}$, $R^{(2)}$ are as in Definition~\ref{defn:def:model:latent_positions:1}.
Since $\MR^{(1)} \cap \MR^{(2)} = \{0 \}$ by definition, and $\MR^{(1)}, \MR^{(2)} \neq \{0\}$ by \refasmp{model:nontriviality:1}, we have a strictly positive separation, that is $\del > 0$. 
 
  We now state our main result.
 
	\begin{thm} \labthm{theorem:error_rate:1}
		Let $A \in \R^{n \times n}$ and $X \in \R^{n \times p}$ be the observed adjacency and covariate matrices as in \refeq{model:sig+noise:1} and let $r_M, r_1, r_2 \in \N$ be the joint and individual subspace dimensions as in \refdef{def:model:latent_positions:1}. Define $(\widehat{M}, \widehat{R}^{(1)}, \widehat{R}^{(2)})$ to be the output of $\refalg{alg:model:spectral:1}$ when supplied $(A, X, r_M, r_1, r_2)$ as arguments. Suppose that \refasmp{model:nontriviality:1} and \refasmp{asmp:estimation:distribution_general} hold, and set   
		\begin{equation}
			\labeq{eq:Estimation Error:epsilons:1}
			\ep^{(1)} = \frac{\kappa \sqrt{r_M+r_1}  }{\lam_{r_M + r_1}(P) },
			\quad \quad  
			\ep^{(2)} = \frac{\tau \sqrt{n(r_M+r_2) (\sig_{r_M + r_2}^2(W) + p ) }}{\sig_{r_M + r_2}^2(W)} \wedge \sqrt{r_M+r_2}
		\end{equation}
		where $\kappa$, $\tau$ are as in \refasmp{asmp:estimation:distribution_general}. Then,  the estimators satisfy
        \begin{align*}
        & \text{Joint components:} \quad & \E[ d(\wh{M}, M)] 
        & = O\bigg(\frac{\sqrt{r_M}}{\del}[\ep^{(1)} +  \ep^{(2)}] \bigg), \\
        & \text{Individual components:} \quad & \E [d(\wh{R}^{(k)}, R^{(k)})] 
        & =  O\bigg(  \frac{\sqrt{r_M r_k}}{\del}[\ep^{(1)} + \ep^{(2)}] \bigg), \hspace{.3cm} k =1,2.
        \end{align*}
	\end{thm}
	
From \refthm{theorem:error_rate:1}, we notice that the error rates are influenced by several factors. 
 First, the terms $\epsilon^{(1)}$ and $\epsilon^{(2)}$ quantify the estimation error for the leading eigenvectors and left singular vectors of the network and the covariates, respectively \citep{Yu:2015:davis_kahan_for_stats,Cai:2018:rate_optim_perturb_bound_sing_subspace}.
 
 Second, the larger is the separation $\delta$ between the individual subspaces $\MR^{(1)}$ and $\MR^{(2)}$, the smaller is the estimation error, matching our intuition.  When $\del$ is close to zero in the presence of noise, the individual components are easily mistaken for joint.

    \refthm{theorem:error_rate:1} provides a general error rate that depends on the spectral norm of the noise in the network adjacency matrix. We now turn our attention to the more concrete setting of independent Bernoulli noise, which has the GRDPG model introduced in Section \ref{Latent Space} as a special case.  To quantify the noise in the Bernoulli model, we denote by $\mu(P)$ the largest expected node degree, that is, 
   $\mu(P) = \max_{i \in [n]} \sum_{j=1}^n P_{ij},$
   and consider the following assumption.
    \begin{asmp} 
    \label{asmp:estimation:distribution:number}
    \labasmp{asmp:estimation:distribution:1}
        Let $A \in \R^{n \times n}$ be the observed adjacency and $P \in \R^{n \times n}$ be its expected value as in \refeq{model:sig+noise:1}. Then, the entries $(A_{ij})_{i \leq j}$ of $A$, are independently distributed with $A_{ij}\sim \text{Ber}(P_{ij})$, and there exists $\rho_n>0$ that only depends on $n$ such that the following conditions hold:
        \begin{enumerate}[label= (\alph*) ]
            \item $\lam_{r_M + r_1}(P) = \Om(\rho_n n)$,
            \item $\rho_n = \om(n^{-1} \log n)$, 
            \item $\mu( P) = \Om (\rho_n n)$.
        \end{enumerate}
    \end{asmp}

    The constant $\rho_n > 0$ is a \emph{sparsity factor} which affects the asymptotic behavior of the largest expected degree $\mu(P)$ and $\lam_{r_M+r_1}(P)$. Condition (b) ensures a lower bound for the rate of the sparsity factor so that under (a), the network signal is sufficiently strong for estimation and under (b) and (c), the network is connected with high probability \citep{Chung:2006:complex_graphs_and_networks}. We now formulate a special case of the above result for this specific Bernoulli model.

	\begin{cor} \labcor{corollary:Estimation Error:consistency:2}
        Let $A \in \R^{n \times n}$ and $X \in \R^{n \times p}$ be the observed adjacency and covariate matrices as in \refeq{model:sig+noise:1} and let $r_M, r_1, r_2 \in \N$ be the joint and individual subspace dimensions as in \refdef{def:model:latent_positions:1}. Define $(\widehat{M}, \widehat{R}^{(1)}, \widehat{R}^{(2)})$ to be the output of \refalg{alg:model:spectral:1} when supplied $(A, X, r_M, r_1, r_2)$ as arguments. Suppose that $\sig_{r_M + r_2}(W) = \Om(\sqrt{np})$, $r_M, r_1, r_2 = O(1)$ and $\del = \Om(1)$. Then under Assumptions \ref{model:nontriviality:number}, \ref{asmp:estimation:distribution_general:number} and \ref{asmp:estimation:distribution:number}
        we have that
        \begin{align*}
         \E[d(\widehat{M}, M)] 
		&=   O\bigg( \frac{1}{\sqrt{\rho_n n}} + \frac{\tau }{\sqrt{p}}  \bigg), \\
         \E[d(\widehat{R}^{(k)}, R^{(k)})] 
		&= O\bigg( \frac{1}{\sqrt{\rho_n n}} + \frac{\tau}{\sqrt{p}}  \bigg),\quad\quad k = 1,2.
        \end{align*}
	\end{cor}

    \refcor{corollary:Estimation Error:consistency:2} provides an explicit characterization of the error rate in terms of the dimensions and magnitude of the noise. Qualitatively, the network sparsity ($\rho_n$) increases the error in the estimation, as well as the variance in the covariate data ($\tau$). The assumptions on the signals $\lam_{r_M+r_1}(P)$ and $\sig_{r_M+r_2}(W)$ are typical for signal-plus-noise and network models \citep{Athreya:2017:survey}.

	\section{Simulations} \label{Simulations}

    In this section, we use simulated data to evaluate the performance of the proposed method for estimating the joint and individual components, as well as compare with alternative approaches. 
    
    \subsection{Example: data with group structure}
    \label{Example: data with group structure}
 We start with an illustration of the decomposition into joint and individual components in latent space models with group structure in the nodes, previously shown in \reffig{fig:sbm:all}.  
We generate a synthetic dataset with $n = 40$ nodes divided into $K = 4$ groups. For simplicity, we enumerate the nodes so that for $ k = 1, \ldots, 4$, nodes $ (k-1) n/4 + 1 \ldots, kn/4$ belong to group $ k$. We demonstrate how the methodology can recover joint and individual group structures in the network and the covariates.

The network data is simulated from a stochastic blockmodel with $K=3$ communities.  Formally, the SBM is defined using a connection probability matrix $B \in [0,1]^{K \times K}$ so that if nodes $i$ and $j$ belong to communities $k$ and $l$ respectively, then the probability of a connection between nodes $i$ and $j$ is 
 $P_{ij} = B_{kl}$. This probability matrix is given by
 $$B = \begin{pmatrix}
     0.6 & 0.05 & 0.05\\
     0.05 & 0.6 & 0.05\\
     0.05 & 0.05 & 0.6
 \end{pmatrix}.$$
 The community labels are encoded in a vector $z^{(1)}\in\{1, \ldots, 4\}^n$, so that $P_{ij} = B_{z^{(1)}_i z^{(1)}_j}$, and we define these memberships as
 $$z_{i}^{(1)} = \left\{\begin{array}{cl}
     1 & i=1, \ldots, 10, \\
     2 & i = 11, \ldots, 20\\
     3 & i = 21, \ldots, 40.
 \end{array}\right.$$
 In this way, the network is only able to identify three of the groups in the nodes, and the nodes from groups 3 and 4 are combined into the same community.

 We consider a similar model for the node covariates by assuming the existence of $L=3$ clusters in a Gaussian mixture model. For each cluster $l = 1, \ldots, L$, there is a corresponding cluster mean $\mu_l \in \R^p$. We assume that for each node $i = 1, \ldots, n$ the expected value of its node covariates, $w_i = \E(x_i)$, satisfies $w_i = \mu_l$ iff node $i$ belongs to cluster $l$. The rows of the covariate matrix are generated from a 3-dimensional Gaussian mixture model, where the cluster means are given by $\mu_1 = 
	\begin{pmatrix}
		-30 & -60 & 30
	\end{pmatrix}^{\top}$,
	$\mu_2 = 
	\begin{pmatrix}
		16 & 8 & 16
	\end{pmatrix}^{\top}$,
	$\mu_3 = 
	\begin{pmatrix}
		-20 & 40 & 20
	\end{pmatrix}^{\top}$. For all clusters, we use the same covariance matrix equal to the identity. 
 The cluster memberships are encoded in a vector $z^{(2)}\in\{1,\ldots, 3\}^n$ in a similar way as in the network, but here the nodes from groups 1 and 2 are combined in the same cluster, resulting in
  $$z_{i}^{(2)} = \left\{\begin{array}{cl}
     1 & i=1, \ldots, 20, \\
     2 & i = 21, \ldots, 30\\
     3 & i = 31, \ldots, 40.
 \end{array}\right.$$
 
 Under this SBM model with covariates, each node $i$ has two different memberships, $z_i^{(1)}$ and $z_i^{(2)}$. Using our model to identify joint and individual components, we distinguish common and individual group structure. 
 In particular, two nodes appear in two different joint clusters if their memberships are different in both datasets, while two nodes appear in different individual clusters if they can be separated by one dataset but not the other. Using \refdef{def:model:latent_positions:1}, the dimensions of the model described above are $r_M = 2$, $r_1 = r_2 = 1$.  \reffig{fig:sbm:joint:2} shows the initial estimates for the joint components following our spectral estimation procedure described previously. The joint components separate the nodes into the two clusters that are common to the network and the covariates. This aligns with the design of `triangle' cluster in network and `green' cluster in the covariates, namely, groups 1-2 and 3-4. \reffig{fig:sbm:indiv:3} shows the estimates for the individual components, each one on a different axis. Looking at the x-axis, the values of $\wh{R}^{(1)}$ are able to distinguish between clusters 1, 2 and 3-4, as this information is individual to the network, whereas the values of $\wh{R}^{(2)}$ in the y-axis can separate groups 3, 4 and 1-2.
 
    \subsection{Numerical comparisons}

    Here we evaluate the performance of our methods and compare with other embedding techniques for network data and covariates.
    \subsubsection{Data generation}

    We consider $n=200$ nodes with $p=10$ covariates for each node and use the model in \refdef{def:model:latent_positions:1} for joint and individual signal components with $r_M=r_1=r_2=1$. We generate  the adjacency matrix $A \sim \operatorname{RDPG}(\sqrt{\alpha} Y)$ of an undirected network with no self-loops (zero diagonal) as in Section \ref{Latent Space}, where  $Y = \begin{pmatrix}
		M & R^{(1)}
	\end{pmatrix} \widetilde{\Gamma}^{(1)}$ is the latent position matrix and $\alpha$ controls the density of the network to make the average expected degree of the nodes equal to $20$.
   For covariates, we generate $X_{ij} \overset{iid}{\sim} N(W_{ij}, \tau^2)$, with $\tau = 0.1$ and
	$W =  \begin{pmatrix}
		M & R^{(2)}
	\end{pmatrix} \Gam^{(2)} $. 
We set
	$$\widetilde{\Gamma}^{(1)} =
	\begin{pmatrix}
		\sqrt{nq_1} & 0 \\
		0 & \sqrt{nq_2}
	\end{pmatrix}, \quad\quad\quad 
	\Gam^{(2)} = 
	\begin{pmatrix}
		\sqrt{ns_1} & 0 \\
		0 & \sqrt{ns_2} 
	\end{pmatrix} Q^{\top}, $$
 where  $q_1, q_2, s_1, s_2 > 0$ are parameters that control the strength of each component,  and $Q \in \O_{p, 2}$ is a random matrix obtained by taking the top two right singular vectors from a $p \times p$ matrix of i.i.d. $N(0, 1)$ entries.

To construct $M$, $R^{(1)}$ and $R^{(2)}$, we consider three orthonormal vectors:
    $$T^{(0)} = \frac{1}{\sqrt{n}}1_n, \quad\quad T^{(1)} = 
    \frac{1}{\sqrt{n}} 1_{n/2} \otimes 
		\begin{pmatrix}
			1 \\
			-1
		\end{pmatrix}, \quad\quad  
    T^{(2)} = \frac{1}{\sqrt{n}} 1_{n/4} \otimes 
		\begin{pmatrix}
			1_2 \\
			-1_2
		\end{pmatrix},$$
  where $1_n$ is the vector of dimension $n$ with all of its entries equal to one. 
We consider two settings in terms of the relative strength of joint and individual signals by defining the components as follows, where $\del = 1 - \l R^{(1)}, R^{(2)} \r$ controls the separation between the individual $R^{(1)}$ and $R^{(2)}$:
	\setlist[description]{font=\normalfont\itshape\textbullet\space}
	\begin{enumerate}[a)]
		\item \textit{Strong joint signal:} $M = T^{(0)}$, $R^{(1)} = T^{(1)}$, $R^{(2)} = \del T^{(1)} + \sqrt{1-\del^2}T^{(2)}$, $q_1 = 0.5$, $q_2 = 0.3$, $s_1 = 0.6$. 
		
	\item \textit{Weak joint signal:} $M = T^{(1)}$, $R^{(1)} = T^{(0)}$, $R^{(2)} = \del T^{(0)} + \sqrt{1-\del^2}T^{(2)}$ and $q_1 = 0.2$, $q_2 = 0.6$, $s_1 = 0.2$. 
	\end{enumerate}

 For each setting, we consider how the performance varies as a function of $\delta$ and as a function of the signal strength of covariates individual component:
 
	\begin{description}
        \item[Varying Individual Subspace Separation:] \ 
		When the joint signal is strong, we set $s_2 = 0.2$, otherwise $s_2 = .7$. We generate the data by varying $\del$ over a grid of $10$ values in $[0,1]$. 
        \item[Varying Individual Component Signal Strength:]\
		We set $\del = 1$ and generate the data by varying $s_2$ over a grid of $10$ values in $[0,1]$. 
	\end{description}

 We consider 50 replications for each combination of parameters and report the average error.


    \subsubsection{Methods for comparison}
    
    We consider: (i) the spectral estimator (\texttt{Spectral}) described in \refalg{alg:model:spectral:1}, (ii) the optimization method from \refalg{alg:model:optimization:2} initialized at the spectral estimate obtained from \refalg{alg:model:spectral:1} (\texttt{Spectral + Optimization}), (iii) the covariate assisted spectral embedding (\texttt{CASE}) from \citep{Binkiewicz:2017:casc}, (iv) the top $r_M$ leading eigenvectors of $A$ (\texttt{Top\_SV\_Net}) and (v) the first $r_M$ left singular vectors of $X$ (\texttt{Top\_SV\_Cov}). Methods (i) and (ii) return estimates of three different components (joint, network individual, and covariate individual), whereas methods (iii)-(v) only return a single set of components. 
 While methods (iii)-(v) do not intend to specifically estimate the joint and individual components of our model, we use them as a baseline since they still capture the true components in the low-rank structure of the data for various parameter settings. We compare the output of each such method to the true joint and individual components separately to investigate under which conditions the components recovered by these methods contain signal specific to the joint or individual components. Explicitly, the components obtained from method (iii) are given by $\sv(L_{\tau}^2 + \al_0 XX^{\top}, 1)$ where $L_{\tau}$ is the regularized graph Laplacian and $\al_0$ is the initial choice of tuning parameter given in \citet{Binkiewicz:2017:casc}. The components obtained from methods (iv) and (v) are given by $\sv(A, 1)$ and $\sv(X, 1)$ respectively. 

To evaluate the performance, we calculate the Procrustes distance from \refeq{eq:procrustes_distance} between the estimated components and the truth separately for joint components $M$ and individual components $R^{(1)}$, $R^{(2)}$. Note that for matrices with orthonormal columns $U, V\in\mathbb{O}_{n,d}$, this distance is bounded by $\sqrt{2d}$, so in our simulations this bound is $\sqrt{2}$.

\subsubsection{Results}
\begin{figure}[!t]
        \includegraphics[width=\textwidth]{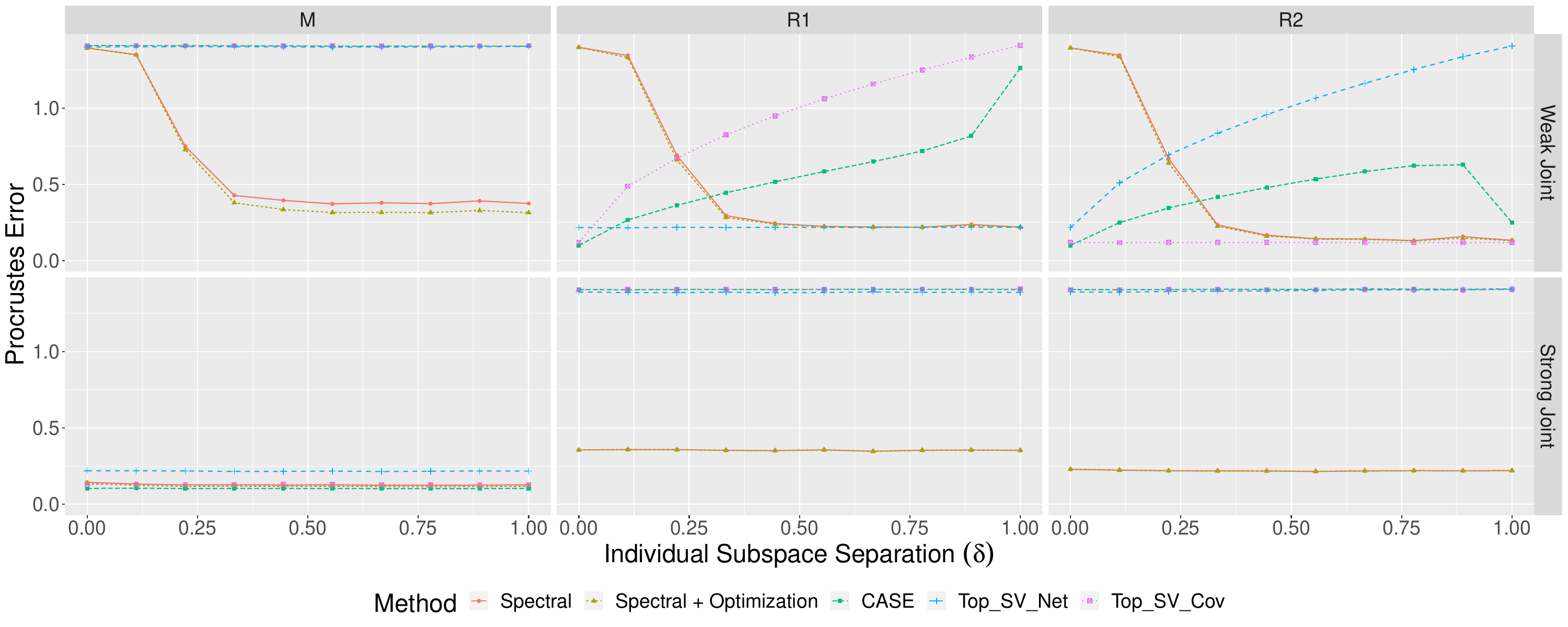}
		\caption{Joint and individual component estimate error (measured by Procrustes distance) averaged over 50 replications as a function of the separation between the individual subspaces ($\del$) while maintaining the individual component signal strength constant.}
		\labfig{fig:simulation:vary_angle:2}
        \label{fig:simulation:vary_angle:2}
	\end{figure}

\begin{figure}[!t]
        \includegraphics[width=\textwidth]{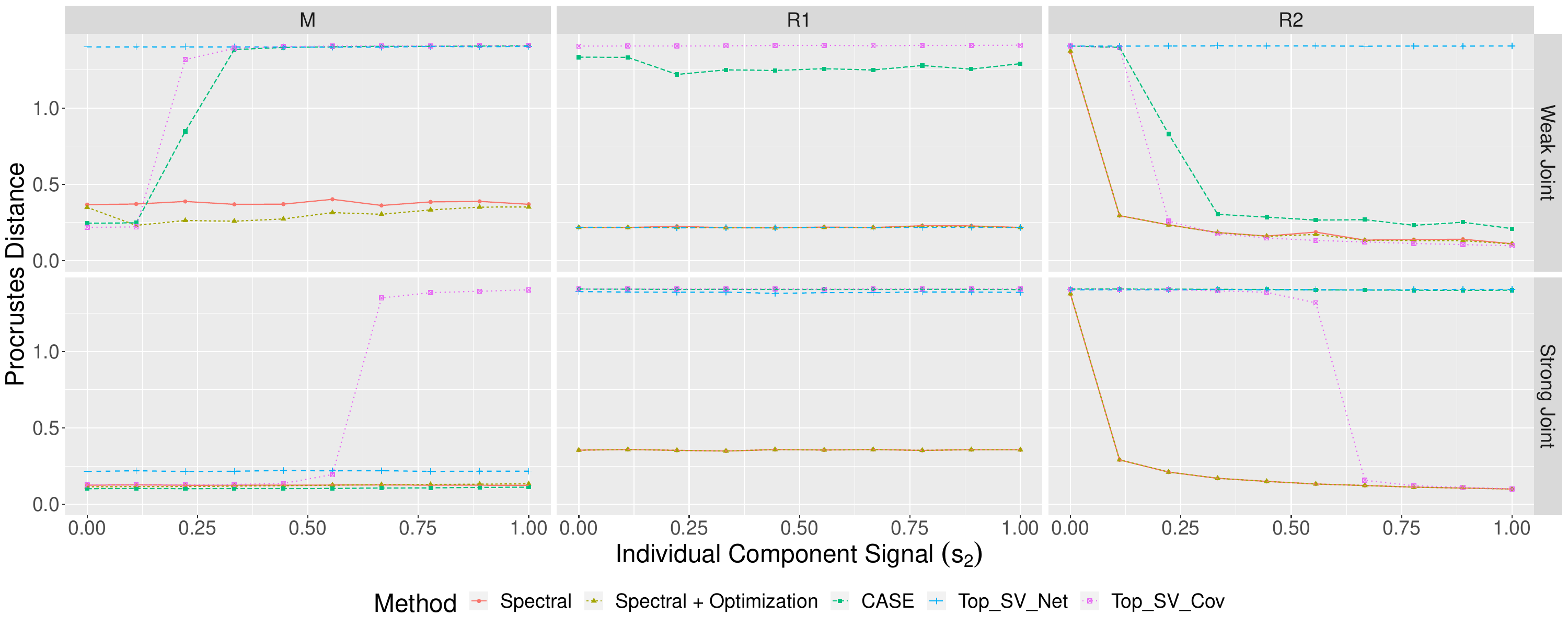}
		\caption{Joint and individual component estimate error (measured by Procrustes distance) averaged over 50 replications as a function of the covariate individual component signal strength $(s_2)$ while maintaining the individual components to be orthogonal.}
		\labfig{fig:simulation:vary_indiv_component_signal:1}
        \label{fig:simulation:vary_indiv_component_signal:1}
	\end{figure}

Figures \ref{fig:simulation:vary_angle:2}-\ref{fig:simulation:vary_indiv_component_signal:1} show the average Procrustes distance over 50 replications for each method, setting and component type. In general, our \texttt{Spectral} and \texttt{Spectral + Optimization} methods show good performance in estimating the correct components when the signal is sufficiently strong. For instance, we observe in \reffig{fig:simulation:vary_angle:2} that the Procrustes error is low whenever the separation between the individual components is sufficiently large (top row) or the signal in the joint components is strong (bottom row).  In \reffig{fig:simulation:vary_indiv_component_signal:1}, our methods perform generally well except for the estimation of the covariate individual components when the signal in this component is low, which is reasonable. These results confirm our observations from \refthm{theorem:error_rate:1} regarding the effect of the $\delta$ and the magnitude of the eigenvalues and singular values of the signal matrices. While both the \texttt{Spectral} and \texttt{Spectral + Optimization} methods show similar performance, we see that in cases where the joint component signal is weak, the optimization method gives an improvement to the results, particularly noticeable in the estimation of $M$.
 
 The alternative approaches (\texttt{CASE}, \texttt{Top\_SV\_Net} and \texttt{Top\_SV\_Cov}) aim to estimate low-rank components that capture the most variability in a given matrix, and as such, they do not differentiate between joint and individual components. Nevertheless, they coincide in some cases with the joint or individual components, which allows us to compare with our methods. When the joint components have strong signal (bottom row of each figure), all of these methods recover $M$ quite well (left column), and our methods show similar performance. On the other hand, when the joint signal is weak (top row of each figure), \texttt{Top\_SV\_Net} and \texttt{Top\_SV\_Cov} estimate the individual components $R^{(1)}$ and $R^{(2)}$, respectively (middle and right columns), and our methods once again show similar performance as long as the subspace separation is large enough. \texttt{CASE} aims to capture the joint signal, but the performance depends on how strong the individual signals are. For instance, \texttt{CASE} generally estimates the individual components when the joint signal is weak (top row of each figure), and the method would either capture $R^{(1)}$ or $R^{(2)}$ depending on which component has the strongest signal. In contrast, our methods can always differentiate between the joint or individual components regardless of the signal strength. 
	
	\section{Application to trade network data with covariates} \label{data_analysis}

 We illustrate the methodology in the analysis of the food trade network between countries reported by the Food and Agriculture Organization of the United Nations (FAO). In addition to the network, we consider country-level economic and development indicators from the World Bank open data database that could explain part of the variation in the trade data. 
 We use our methodology to explore the joint and individual structure in the trade data and the covariates, with the goal of understanding trading patterns that can be explained by country-level indicators, as well as additional information that is unique to each dataset.
 
	\noindent\tbf{Data description and processing.}
	The version of the FAO trade data we used was collected by \cite{domenico_trade_data} and represents the total food trade from one country to another of a given food product during the year 2010. For the node covariate data, we collected economic and development country data from the same year reported in the World Bank open data database through the \texttt{wbstats} package \citep{wbstats}.
	
	The original network data reports multiplex-directed food trade between countries. Each of the 364 layers of the multiplex network corresponds to a food product. Nodes correspond to countries and edges correspond to the import and export relations between them for the associated food product, weighted by the traded quantity of the particular product. We aggregate the adjacency matrices over all layers to obtain a single weighted and directed adjacency matrix representing the total food import/export trade between countries. We then symmetrize this adjacency matrix by adding its transpose to itself. Finally, the weights are log-transformed by $\log(x+1)$.
	
	The covariate data consists of the following economic and developmental indicators: (i) gross domestic product per capita in US dollars (GDP), (ii) agricultural, forestry and fishing value added as a percentage of the GDP (AFF), (iii) total natural resource rents as a percentage of the GDP (NAT), (iv) government expenditure on education as a percentage of the GDP (EDU) and (v) crude birth rate per 1000 people (BIR). The values of these covariates were log-transformed similarly to the trade network weights. In addition, categorical covariates indicating the continents to which each country belongs (Africa, Europe, the Americas, Asia and Oceania) are represented via additional columns in the covariate matrix with dummy variables. 
    The columns of the covariate matrix were then centered and standardized.
    
    As the countries with available data from the FAO data are not the same as those with available data from the World Bank dataset, we only consider the 146 countries with data present in both. In summary, we work with a symmetric, weighted adjacency matrix $A \in \R^{146 \times 146}$ and a covariate matrix $X \in \R^{146 \times 10}$.

    \noindent\tbf{Data analysis.}
    We begin with an exploratory analysis of each dataset independently by constructing separate node embeddings for the network and the covariates. For the network, we compute the adjacency spectral embedding (ASE) \citep{Sussman:2012:consistent_ase} into four dimensions, which is defined as $\widehat{Y}=\widehat{V}^{(1)}|\widehat{\Lambda}|^{1/2}$, where $\widehat{V}^{(1)} = \eig(A, 4)$ is the matrix of leading eigenvectors of $A$ and $\Lambda  = \text{diag}(\lambda_1(A), \ldots, \lambda_4(A))$ are its leading eigenvalues. The rows of the ASE are a consistent estimator for the latent positions of the GRDPG model defined in Section~\ref{Latent Space} \citep{Rubin-Delanchy:2022:GRDPG}.
    For the covariate data, we obtain the top four left leading singular vectors, which are equivalent to the scores of the principal components. Ideally, any nontrivial structure should be identified by these components as they capture most of the variation in the data. We display these plots in \reffig{fig:data:ASE_net_cov}. Upon inspection, the network embeddings in \reffig{fig:data:ASE_net:13} only display part of the geographic structure related to the region (continents). In contrast, the covariate principal components display a clustering structure according to continent in \reffig{fig:data:ASE_cov:14}, as these indicator variables are included in the covariate data. The PCA biplot shows two main directions of variability corresponding to the size of the economy (GDP per capita and other variables correlated to it), and the expenditure in education (EDU). 

    We now estimate joint and individual components from the data. 
    To estimate the ranks in the model $\rnk(P) = r_M + r_1$ and $\rnk(W) = r_M + r_2$, we first estimate the embedding dimensions via the elbow method on the scree plots of the singular values of the adjacency and covariate matrices, resulting in values $\rnk(P) = 4$ and $\rnk(W) = 4$. For the dimensions of the joint and individual components, we first estimate the dimension of the joint structure based on the scree plot of the singular values of the matrix $\wh{U}$ obtained in \refalg{alg:model:spectral:1} and look for an elbow. This results in an estimated value of $r_M=2$, and therefore, we set $r_1 = 2$ and $r_2 = 2$.

    \begin{figure}[!t] 
        \centering
        \begin{subfigure}[]{.475\textwidth}
			\includegraphics[scale = .5]{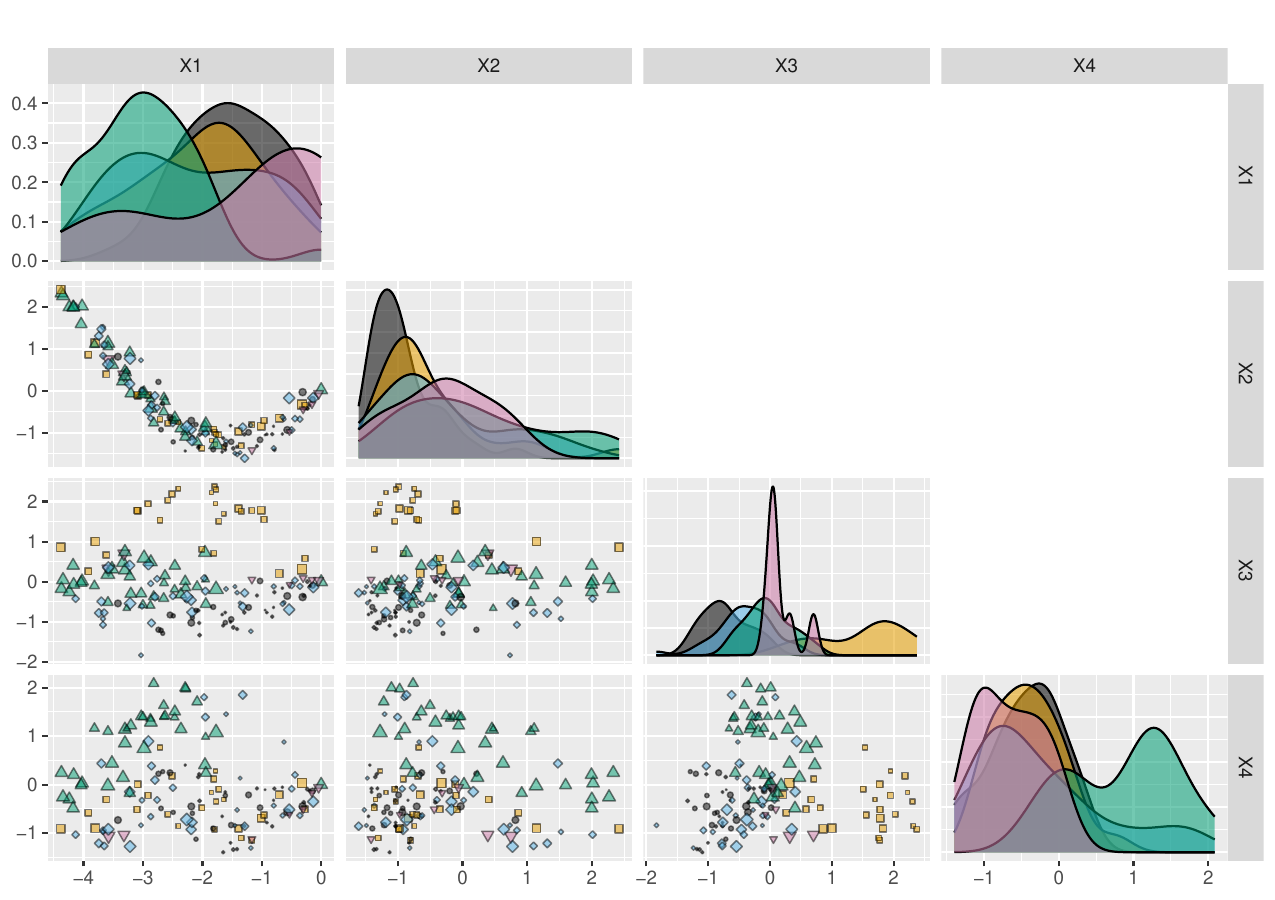}
			\caption{ASE of the trade network}
            \labfig{fig:data:ASE_net:13}
		\end{subfigure}
        \hfill
        \begin{subfigure}[]{.475\textwidth}
			\includegraphics[trim={3.5cm 0cm 0cm 0cm}, clip, scale = .5]{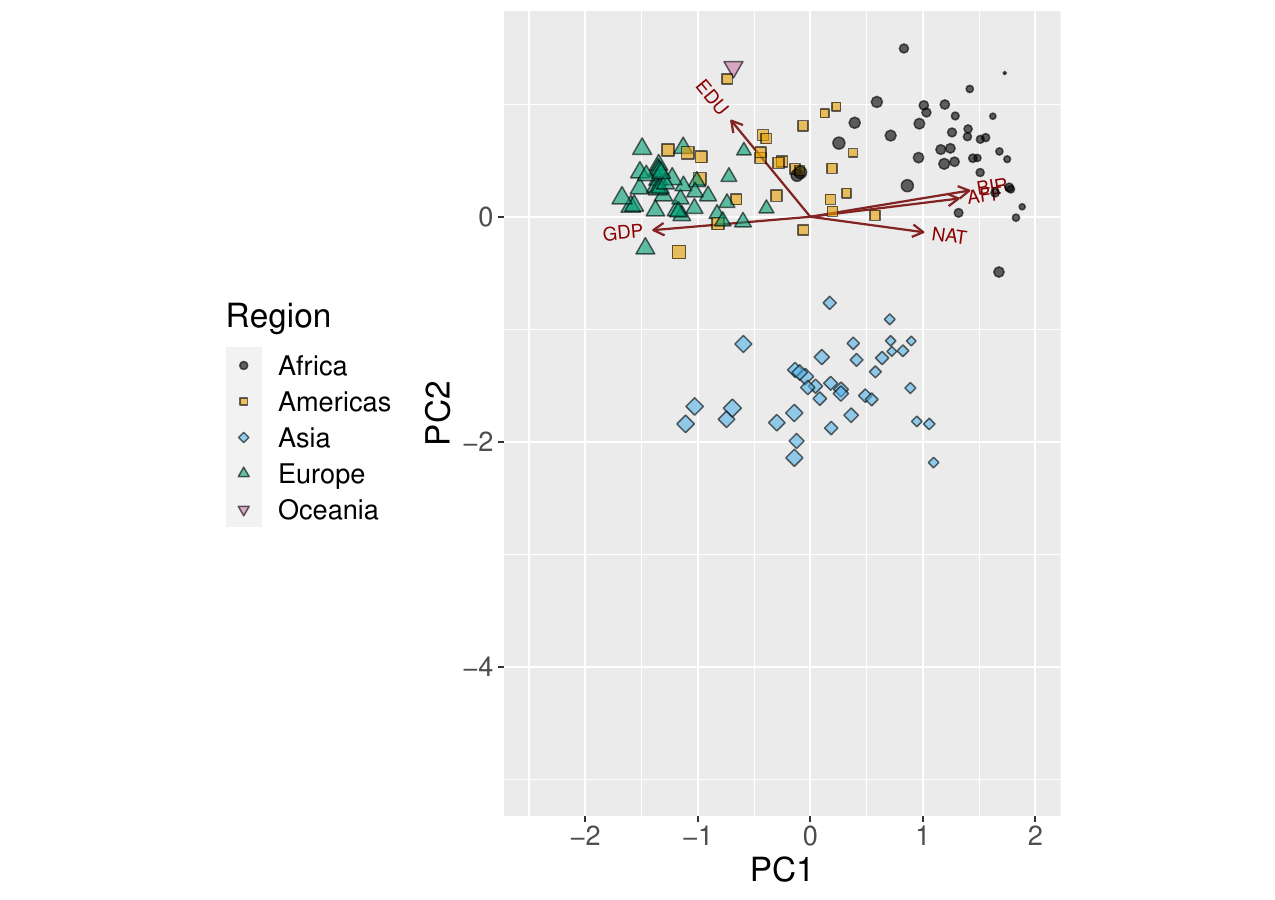}
			\caption{Biplot of the covariate principal components}
            \labfig{fig:data:ASE_cov:14}
		\end{subfigure}
		\caption{Network and covariate embeddings, constructed by separately computing the network adjacency spectral embedding (ASE) and the covariate principal components. Each point represents a different country (node), colored by continent and size proportional to GDP per capita.}
		\labfig{fig:data:ASE_net_cov}
	\end{figure}

 The joint and individual components are first estimated with the spectral method (\refalg{alg:model:spectral:1}), and these are used to initialize the optimization method (\refalg{alg:model:optimization:2}). The joint components obtained by each of these methods are shown in Figures
 \ref{fig:fig:data:joint_spectral:5} and \ref{fig:fig:data:joint_optim:6}. While both methods show a very similar pattern, the joint components from the optimization method show tighter groupings than those obtained from the spectral method. We observed a similar result in the individual components, and hence, we only report the results for the network and covariate individual components obtained from the optimization method in~\reffig{fig:data:indiv_optim:10}.

    \begin{figure}[!t] 
        \centering
		\begin{subfigure}[]{.45\textwidth}
			\includegraphics[scale = .375]{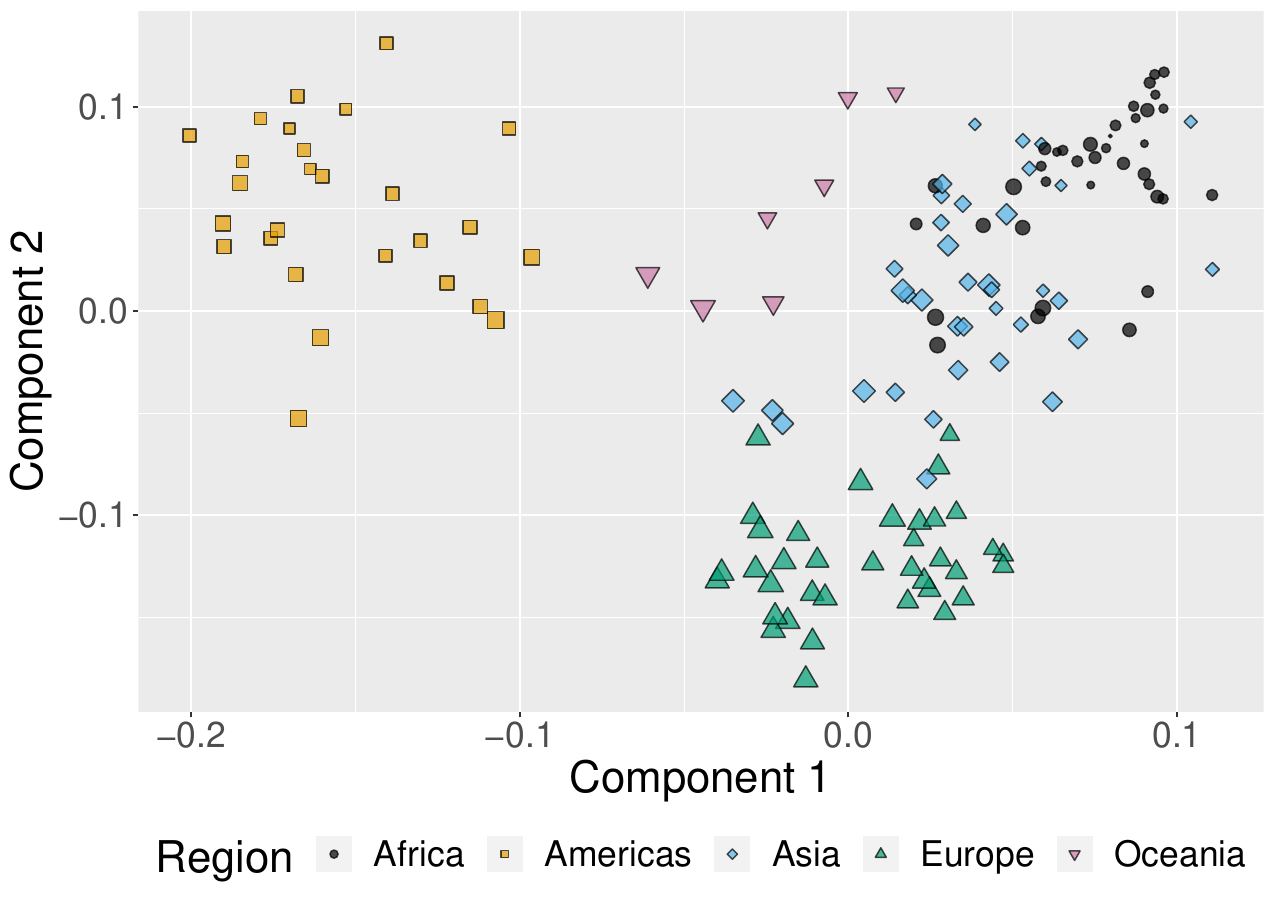}
			\caption{Spectral method}
            \labfig{fig:data:joint_spectral:5}
		\end{subfigure}
        \hspace{1em}%
		\begin{subfigure}[]{.45\textwidth}
			\includegraphics[scale = .375]{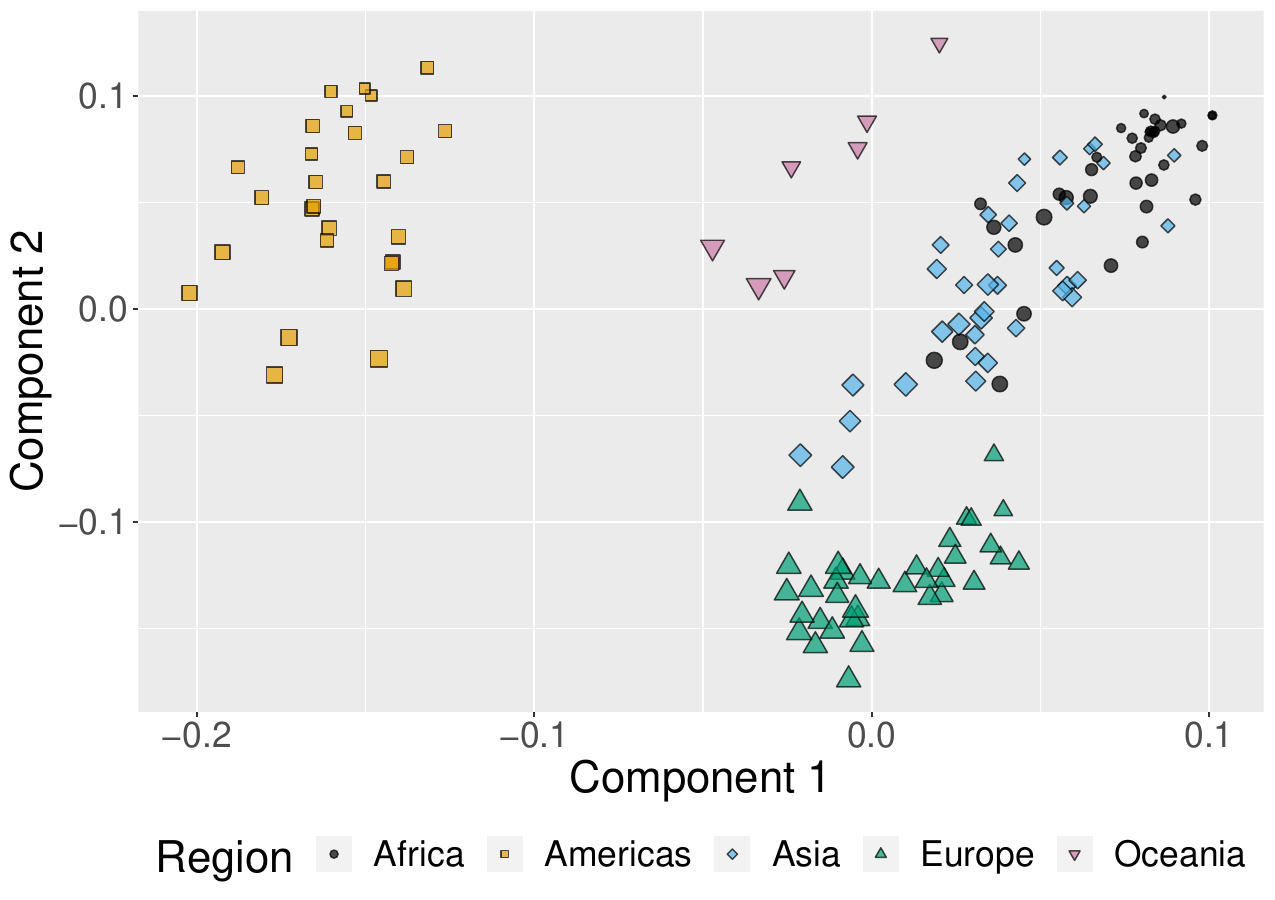}
			\caption{Optimization method}
			\labfig{fig:data:joint_optim:6}
		\end{subfigure}
		\caption{Plots of estimated joint components ($\widehat{M}$) from \refalg{alg:model:spectral:1}  (left) and \refalg{alg:model:optimization:2}  (right). Each point represents a different country (node), colored by continent and size corresponding to GDP value.}
    \labfig{fig:data:joint_spectral_optim:7}
	\end{figure}

 \begin{figure}[!t] 
		\centering
		\begin{subfigure}[]{0.45\textwidth}
			\includegraphics[scale = .375]{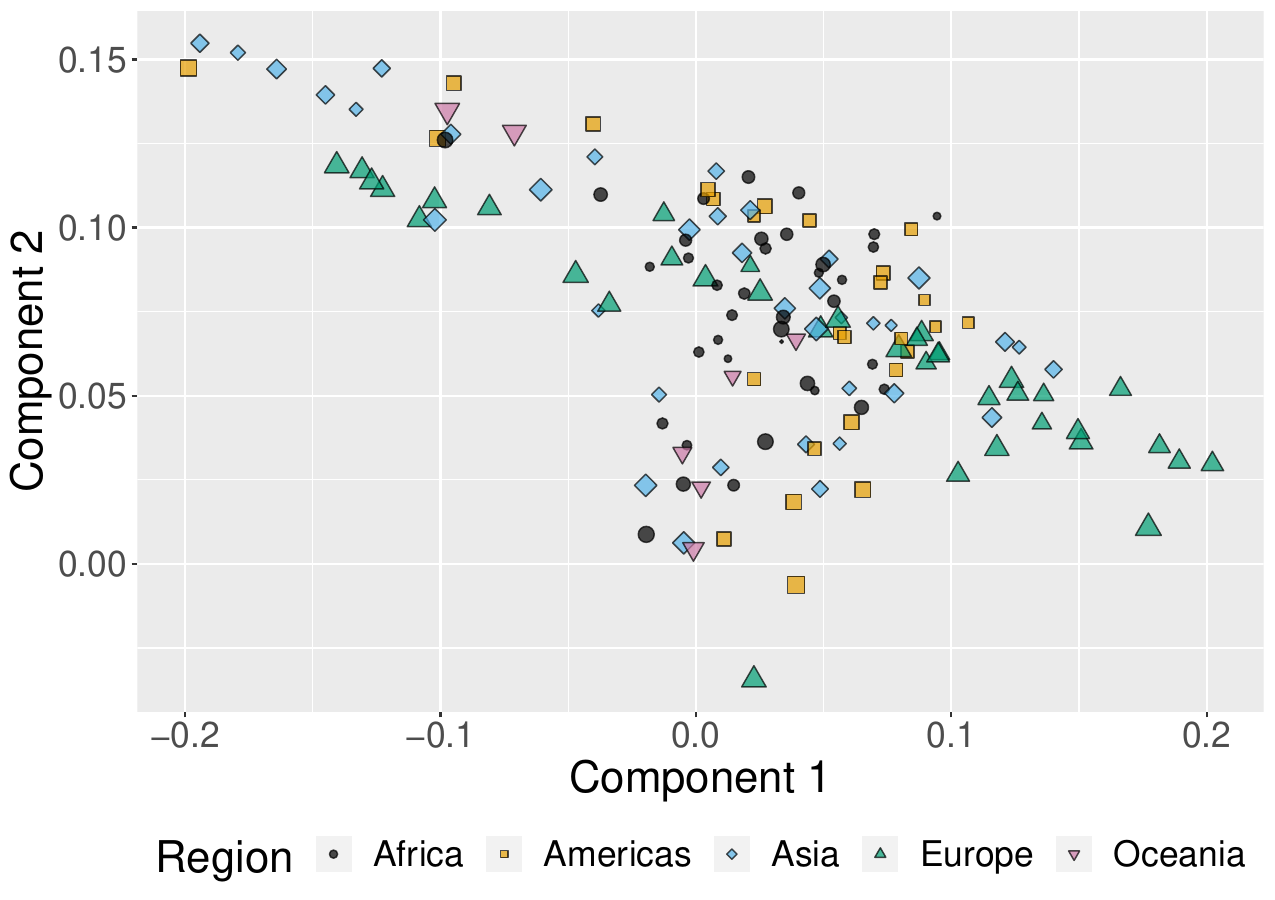}
			\caption{Network individual components}
            \labfig{fig:data:network_indiv_optim:8}
		\end{subfigure}
        \hspace{1em}%
		\begin{subfigure}[]{0.45\textwidth}
			\includegraphics[scale = .375]{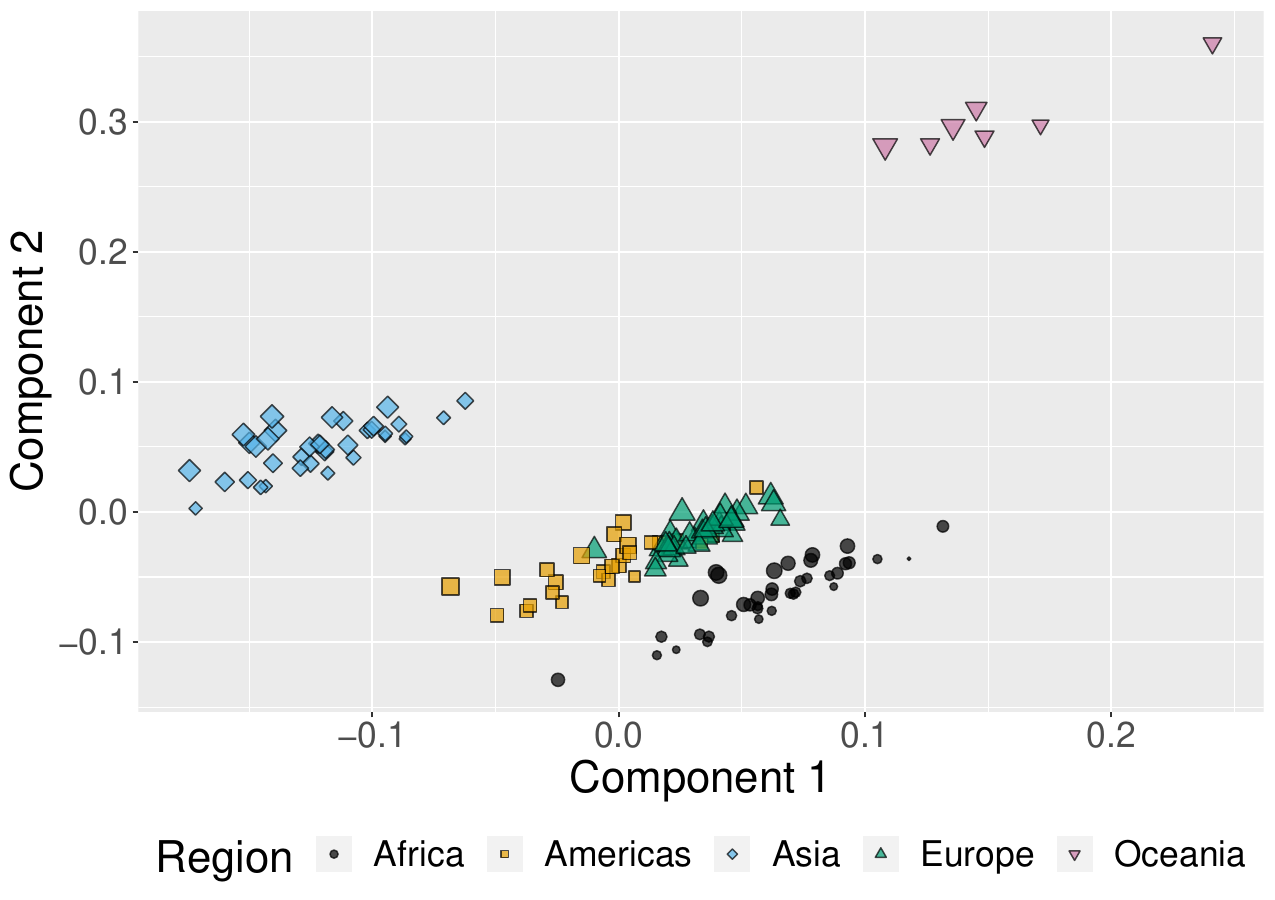}
			\caption{Covariate individual components}
   \labfig{fig:data:covar_indiv_optim:9}
		\end{subfigure}
		\caption{Plots of estimated individual components for the network ($\widehat{R}^{(1)}$) and covariates ($\widehat{R}^{(2)}$) using \refalg{alg:model:optimization:2}.}
    \labfig{fig:data:indiv_optim:10}
	\end{figure}

 \begin{figure}[!t] 
    \centering
    \begin{subfigure}[]{0.45\textwidth}
	\includegraphics[scale = .375]{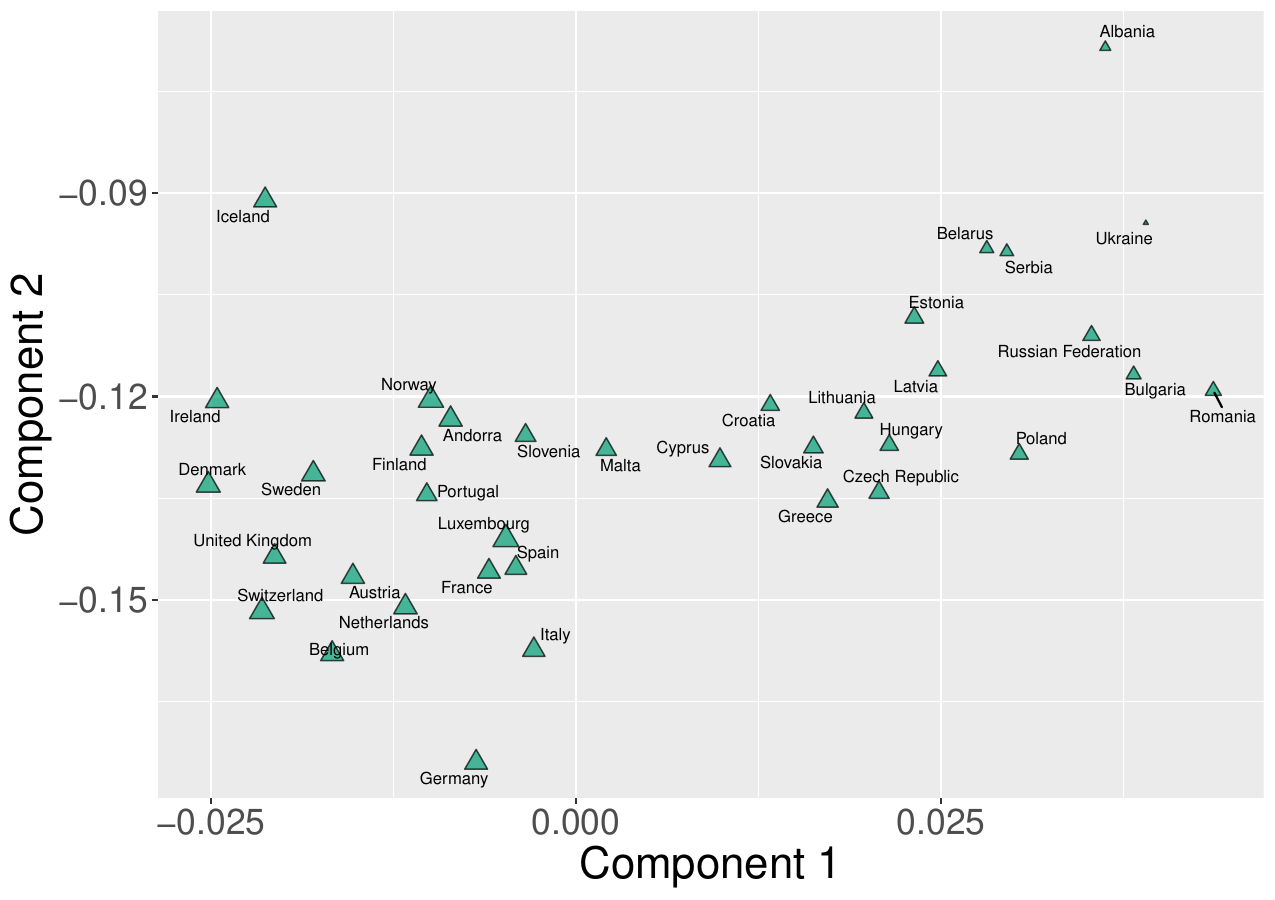}
	\caption{Europe}
        \labfig{fig:data:joint_optim:europe}
    \end{subfigure}
    \hspace{1em}%
    \begin{subfigure}[]{0.45\textwidth}
	\includegraphics[scale = .375]{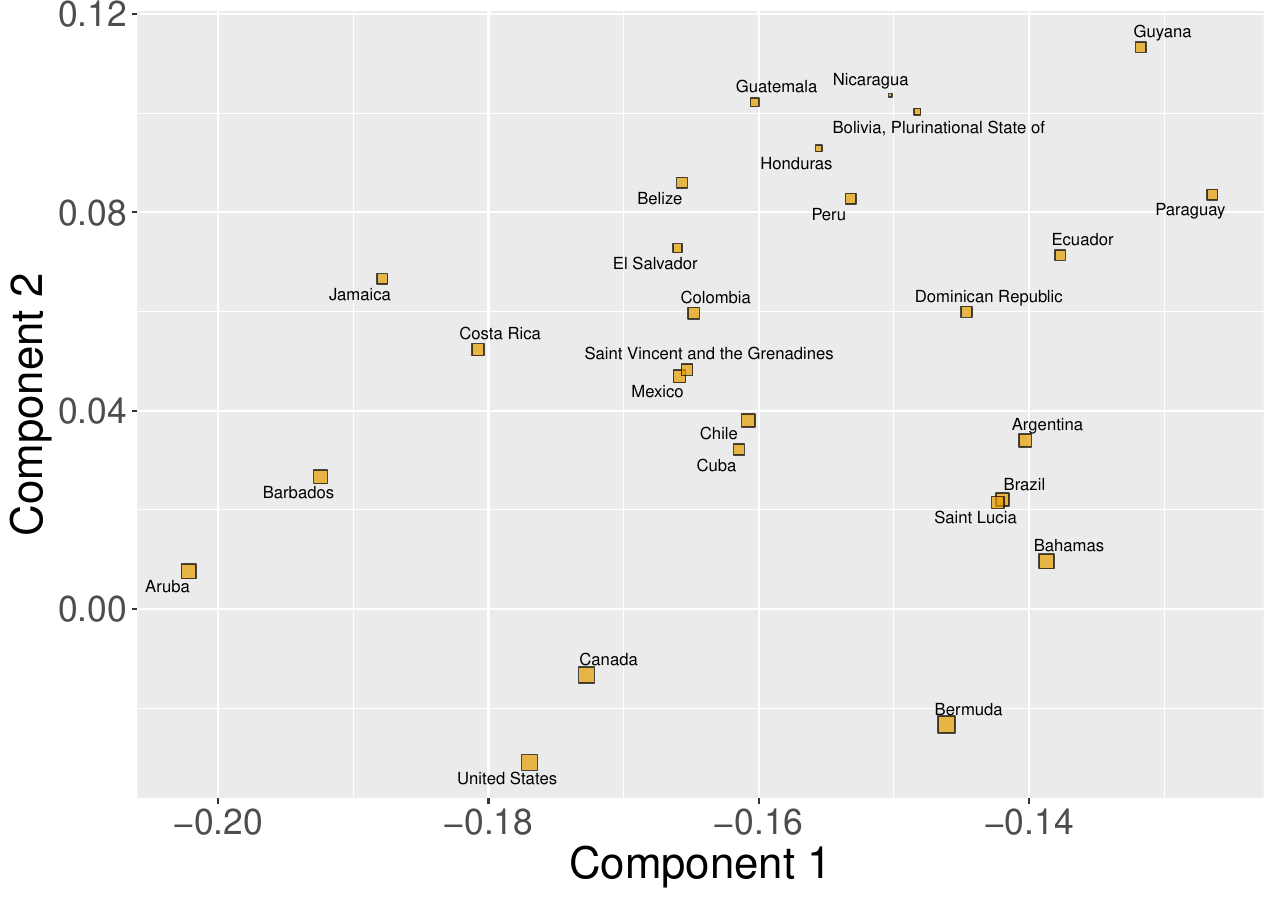}
	\caption{Americas}
        \labfig{fig:data:joint_optim:americas}
    \end{subfigure}
    \caption{Estimated joint components for countries in Europe and the Americas. Each point size is proportional to the country GDP per capita.}
    \label{fig:data:joint_europe_and_americas}
\end{figure}

\begin{figure}[!t] 
    \centering
    \begin{subfigure}[]{0.45\textwidth}
        \includegraphics[scale = .375]{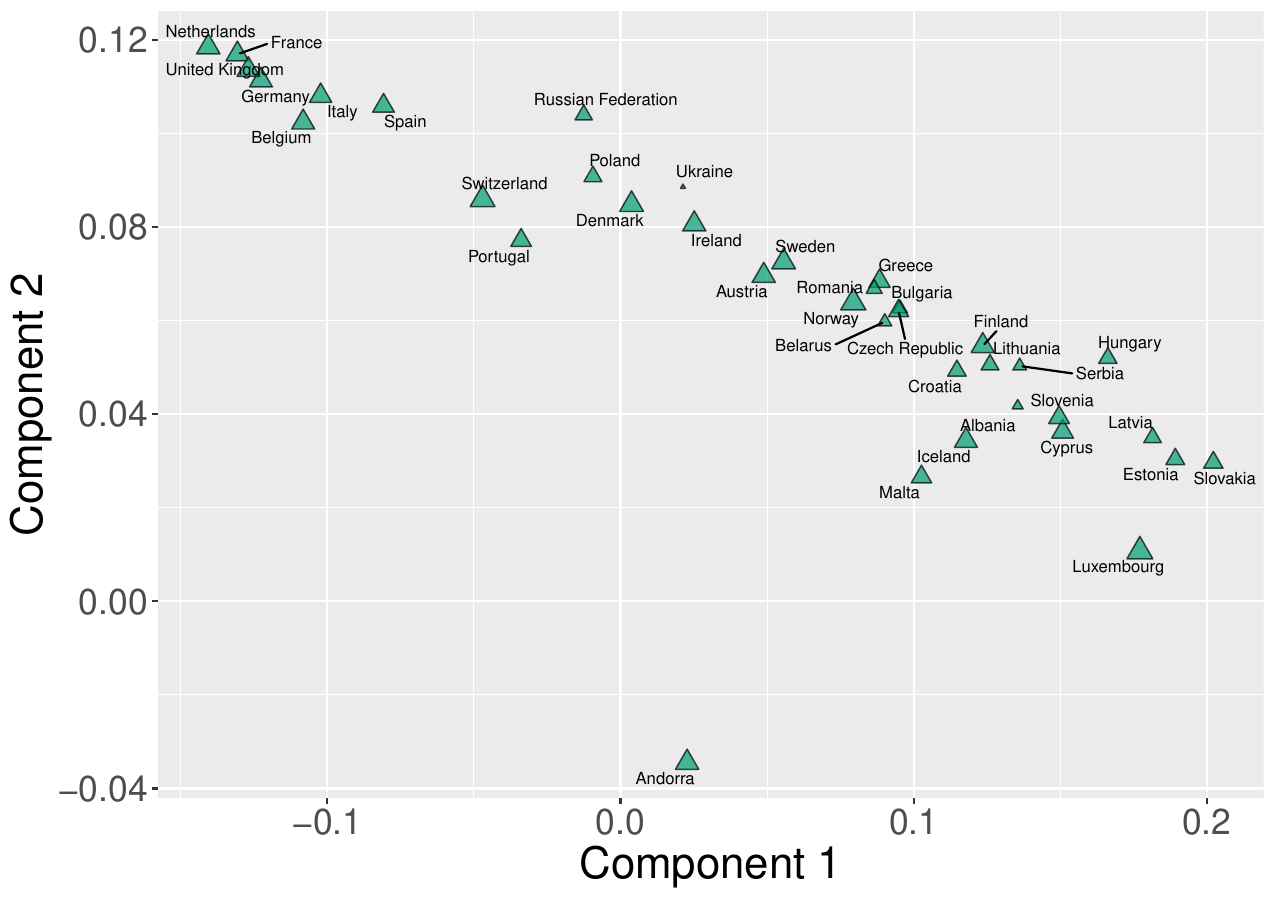}
        \caption{Europe}
        \labfig{fig:data:network_indiv_optim_europe}
    \end{subfigure}
    \hspace{1em}%
    \begin{subfigure}[]{0.45\textwidth}
        \includegraphics[scale = .375]{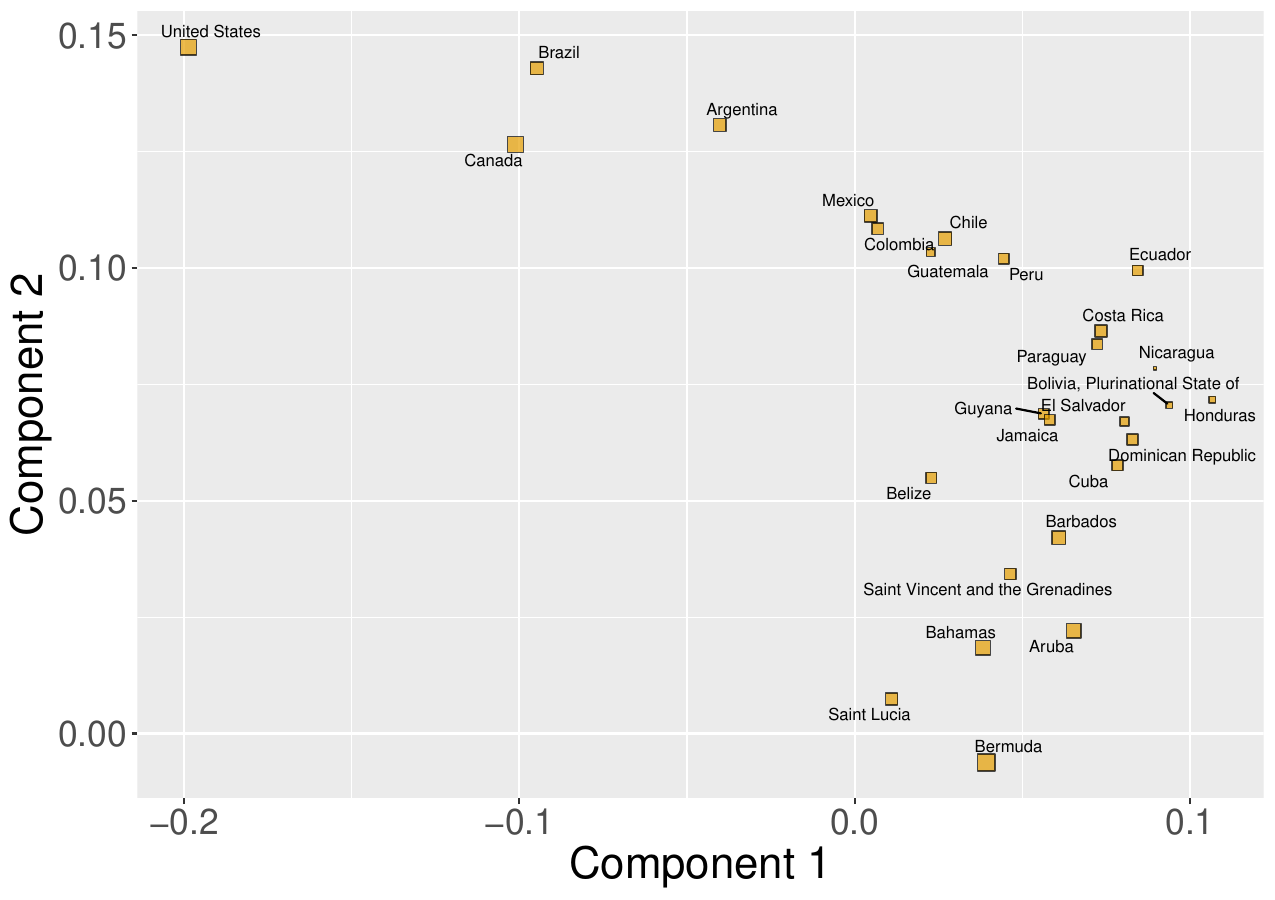}
        \caption{Americas}
        \labfig{fig:data:network_indiv_optim_americas}
    \end{subfigure}
    \caption{Estimated network individual components for countries in Europe and the Americas. Each point size is proportional to the country GDP per capita.}
    \label{fig:data:americas_europe_net_indiv}
\end{figure}

 \begin{figure}[!t] 
    \centering
    \begin{subfigure}[]{0.45\textwidth}
         \centering
        \includegraphics[scale = .375]{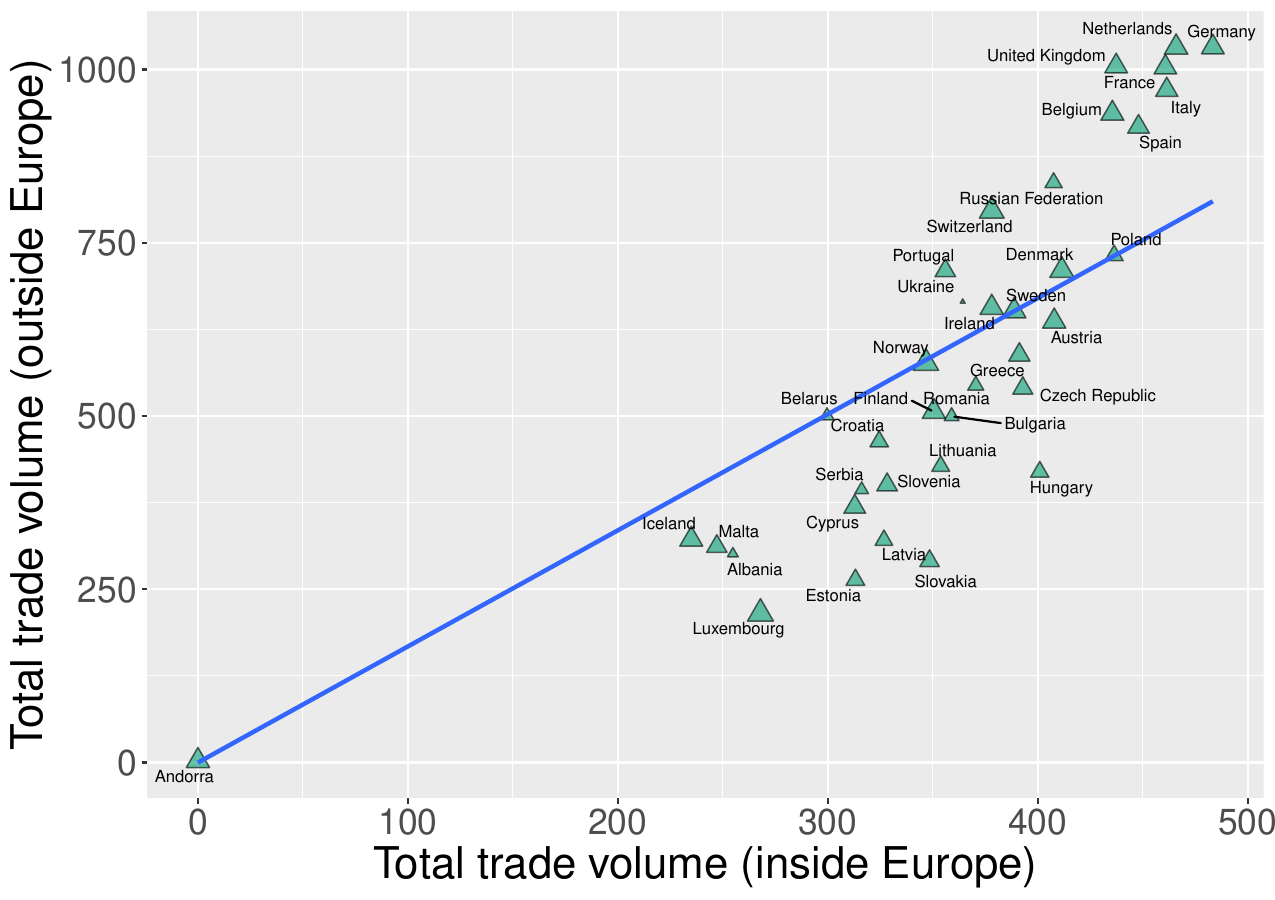}
        \caption{total trade inside vs total trade outside Europe for European countries.}
        \labfig{fig:data:inside_outside_trade_volume_europe}
    \end{subfigure}
   \hspace{1em}%
    \begin{subfigure}[]{0.45\textwidth}
        \centering
        \includegraphics[scale = .375]{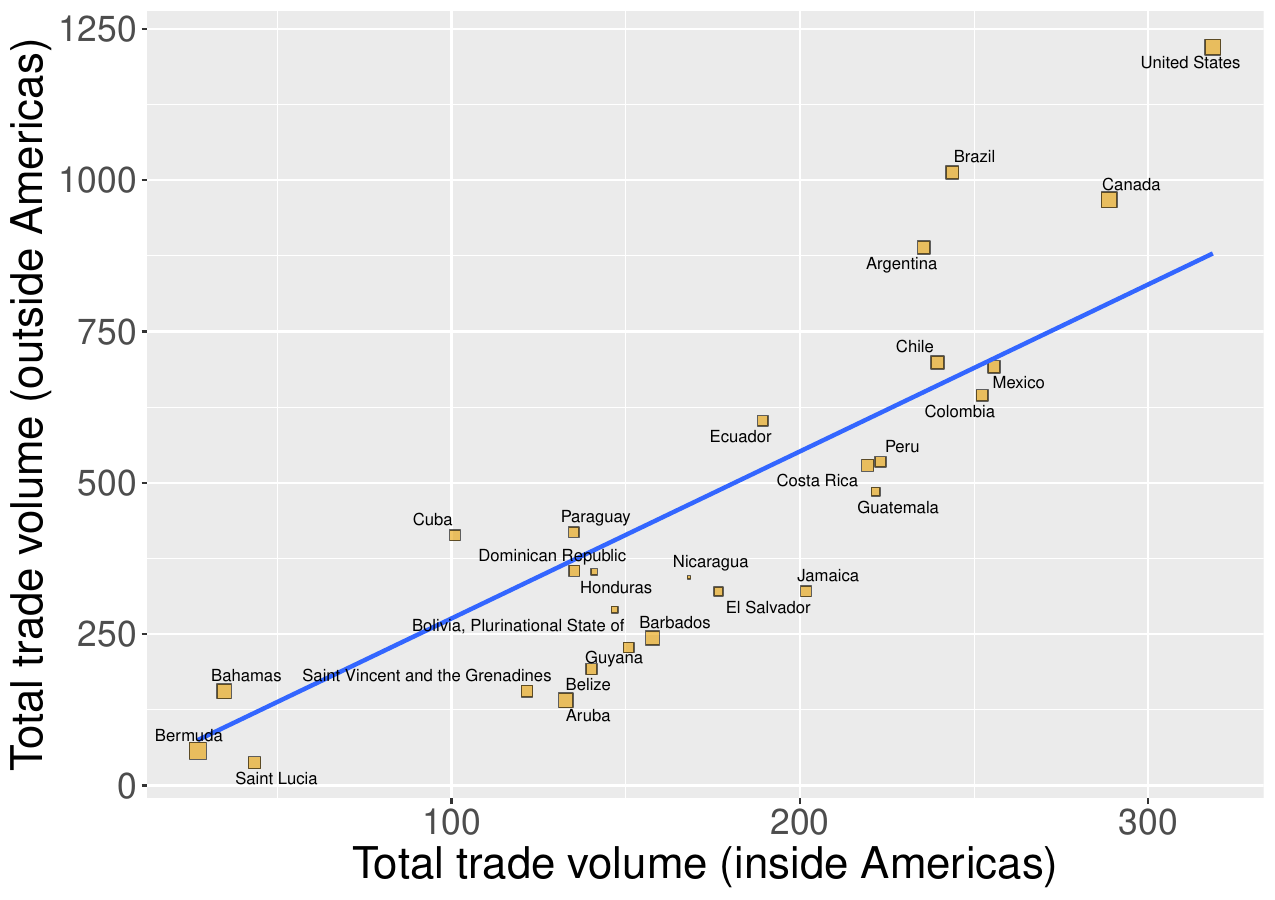}
        \caption{total trade inside vs total trade outside the Americas for American countries.}
        \labfig{fig:data:inside_outside_trade_volume_americas}   
    \end{subfigure}
    \caption{Total trade volume per country divided into trade with other countries inside (x-axis) and outside (y-axis) the same region. The blue line corresponds to the fitted regression value of a linear model without an intercept; countries below this line have higher local trade volume than average and vice-versa. The location of the countries in this figure closely resembles the relative location in the individual network components (Figure \ref{fig:data:americas_europe_net_indiv}).}
    \labfig{fig:data:inside_outside_trade_volume}
\end{figure}

To gain more insight into the information displayed in the estimated components, we focus on the countries from specific regions, in particular, the {Americas} and {Europe}. Figures \ref{fig:data:joint_europe_and_americas} and \ref{fig:data:americas_europe_net_indiv} show the estimated joint and individual components only for the countries in those regions, together with the country names and the point size proportional to the GDP per capita. In the joint components (Figure \ref{fig:data:joint_europe_and_americas}), the scatterplots show an association between GPD per capita and the joint latent positions, as countries with higher GDP are located closer to the bottom left of the corresponding plots. This observation suggests that this covariate plays an important role in explaining trade between countries. The individual components, on the other hand, do not capture a clear association with GDP, as this was already captured in the joint. 

The individual components in Figure \ref{fig:data:americas_europe_net_indiv} reveal trading structure not captured by the covariates. A further analysis of the data reveals that the components capture global and local trade variability, as countries located closer to the right tend to trade more with countries from the same region, whereas countries on the left side have higher trade volume with countries from a different region. This trend is confirmed in \reffig{fig:data:inside_outside_trade_volume}, which shows the total trade volume of each country with other countries from the same region against the total trade volume with countries from different regions. For each country $i$, these quantities are defined as
$$w^{\text{inside}}_i = \sum_{j: \text{region}(i) = \text{region}(j)} A_{ij}, \quad\quad\quad\quad w^{\text{outside}}_i = \sum_{j: \text{region}(i) \neq \text{region}(j)} A_{ij}.  $$
Comparing  \reffig{fig:data:network_indiv_optim_americas} and \reffig{fig:data:network_indiv_optim_europe}, we see that the network individual components are capturing much of the same information as the total trade volume inside and outside of the respective region. Moreover, this information is not associated with GDP per capita, as countries with high GDP appear on both sides of the figure.

	To gain more interpretation of the role of the covariates in explaining the components, we decompose the covariate matrix by projecting the data onto the joint and individual subspaces as
 \begin{equation}
 \labeq{eq:data_analysis:decomp}
X = \MP_{\wh{M}}X + \MP_{\wh{R}^{(2)}}X + (I - \MP_{\wh{M}} -\MP_{\wh{R}^{(2)}})X.
 \end{equation}
  For the joint and individual projections, we do PCA to obtain the loadings of each of the covariates onto the joint and individual components. Figures~\ref{fig:fig:data:covar_joint_PCA:10} and \ref{fig:fig:data:covar_indiv_PCA:11} show the biplots representing the country scores obtained from each of the projections, together with the variable loadings (arrows in the figure). 
  We see in \reffig{fig:data:covar_joint_PCA:10} that there is a large correlation between the economic covariates (GDP, BIR, AFF) and the first principal component in the joint PCA, suggesting that the effect captured by both the network and covariate data is related to these factors. This observation coincides with the results in Figure~\ref{fig:data:joint_europe_and_americas}. On the other hand, the loadings for education are low in the joint components, which is reasonable given that we do not expect that government expenditure on education would impact the trading network. The projection of the covariate data onto the individual components captures the remaining variability with respect to the economic factors, and the loadings for education and natural resources show higher values, indicating that these variables do not have a strong impact on the network trading patterns.
	
	\begin{figure}[!t] 
		\centering
		\begin{subfigure}[b]{0.45\textwidth}
        \includegraphics[trim={3.75cm 0cm 0cm 0cm}, clip, scale = .5]{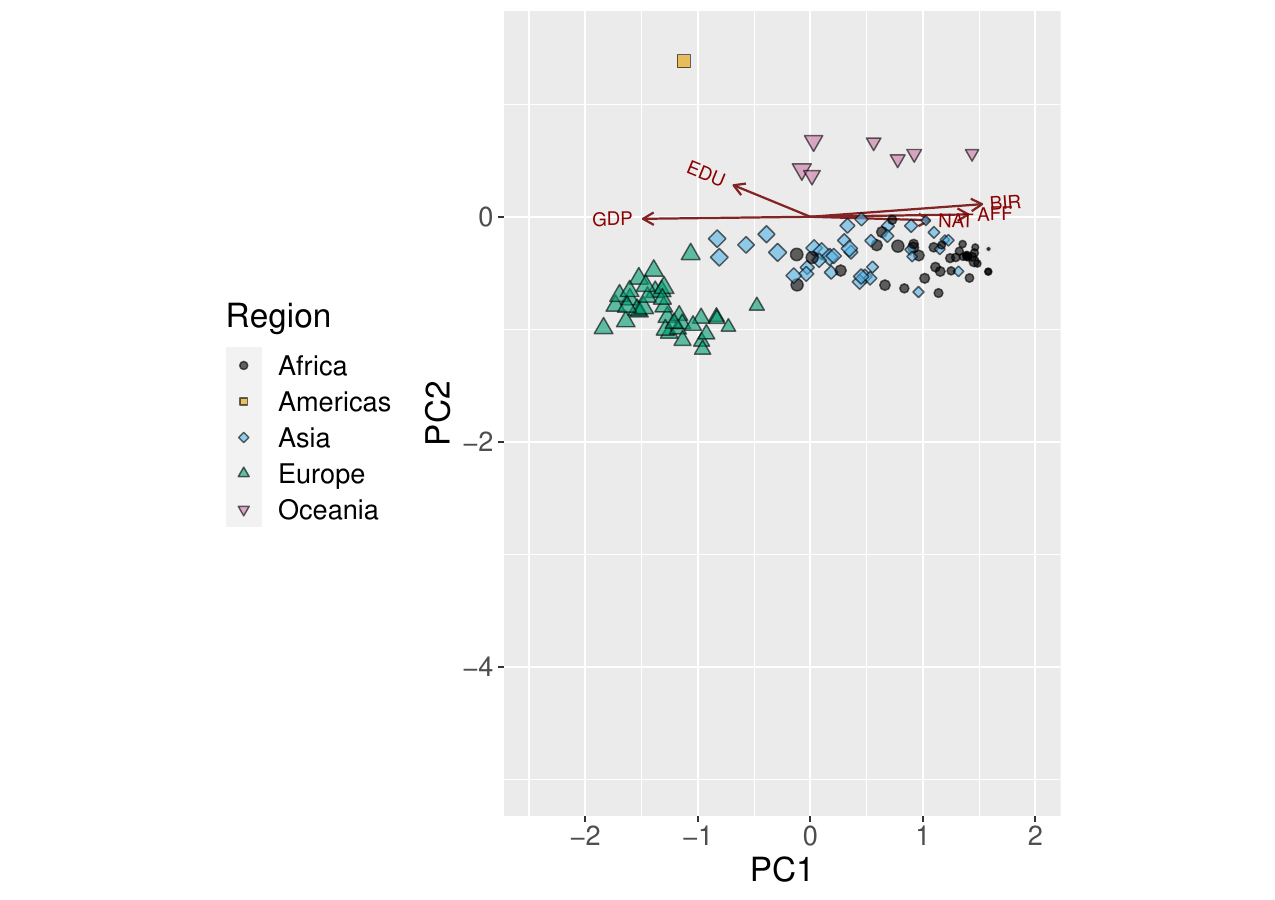}
			\caption{PCA biplot for $\MP_{\wh{M}}X$}
			\labfig{fig:data:covar_joint_PCA:10}
		\end{subfigure}
		\begin{subfigure}[b]{0.45\textwidth}
			\includegraphics[trim={7cm 0cm 2cm 0cm}, clip, scale = .5]{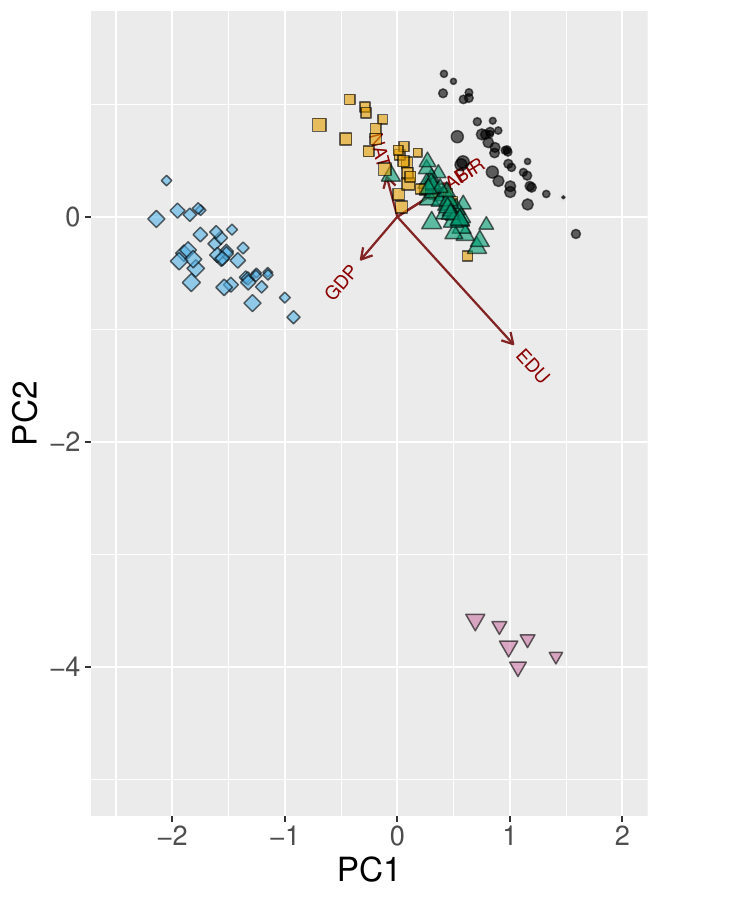}
			\caption{PCA biplot for $\MP_{\wh{R}^{(2)}}X$}
			\labfig{fig:data:covar_indiv_PCA:11}
		\end{subfigure}
		\caption{PCA biplots for the joint and individual parts of the covariates}
	\end{figure}

   Finally, we measure the proportion of variation explained by the joint and individual components. For the covariates, we use the decomposition in \refeq{eq:data_analysis:decomp} to calculate the proportion of variation in the covariates explained by the joint, individual and residual components, denoted $\Var_J(X)$, $\Var_I(X)$ and $\text{Var}_{\text{res}}(X)$ respectively, as 
    $$\text{Var}_{J}(X) =\frac{\|\MP_{\MC(\widehat{M})} X \|_F^2} {\| X \|_F^2}, \quad \quad  \text{Var}_{I}(X) =\frac{\|\MP_{\MC(\widehat{R}^{(2)})} X \|_F^2}{\|X\|_F^2}, \quad \quad \text{Var}_{\text{res}}(X) = 1 - (\text{Var}_{J}(X) + \text{Var}_{I}(X)).$$
    For the network adjacency matrix, we first decompose the variation into parts explained by the signal and the residual as
    $$A = \widehat{P} + (A - \widehat{P}),$$
    where $\widehat{P}= \MP_{(\wh{M}, \wh{R}^{(1)})} A\MP_{(\wh{M}, \wh{R}^{(1)})}^\top$ is the projection of $A$ onto the joint and individual components. We denote the proportion of signal and residual variation as $\text{Var}_{\wh{P}}(A)$ and $\text{Var}_{\text{res}}(A)$ respectively, defined by  
    $$\text{Var}_{\wh{P}}(A) =  \frac{\|\wh{P}\|_F^2}{\| A \|_F^2}, \quad \quad \text{Var}_{\text{res}}(A) =  1 - \text{Var}_{\Sig}(A).$$
    We then define the proportion of variation in the network signal explained by the joint and individual components in terms of the latent positions of the model as in \refeq{eq:model:latent_pos}, denoted $\text{Var}_{J|\wh{P}}(A)$ and $\text{Var}_{I|\wh{P}}(A)$, respectively. Specifically, let $\wh{Y} = \operatorname{ASE}(\wh{P}, 4)$ be an estimate of the latent positions of the model. We set 
    $$ \quad \text{Var}_{J|\wh{P}}(A) = \frac{\| \MP_{\widehat{M}} \wh{Y} \|^2_F}{\| \wh{Y} \|^2_F},\quad\quad\quad \quad \text{Var}_{I|\wh{P}}(A) = \frac{\| \MP_{\widehat{R}^{(1)}} \wh{Y} \|^2_F}{\| \wh{Y} \|^2_F}$$ 
    respectively. We then define the proportion of variation in the network explained by the joint and individual components, denoted $\text{Var}_J(A)$ and $\text{Var}_J(A)$ respectively, by 
    $$\text{Var}_J(A) = \text{Var}_{J|\wh{P}}(A) \text{Var}_{\wh{P}}(A), \quad \quad \text{Var}_I(A) = \text{Var}_{I|\wh{P}}(A) \text{Var}_{\wh{P}}(A). $$
    The proportions of variation explained are reported in \reftable{fig:table:variation_explained}. Overall, the low-rank models capture a large portion of the variation in the data (88\% for the network and 79.75\% for the covariates). However, when looking at the joint and individual components, the proportions vary significantly in the network and the covariates. For instance, while most of the variation in the covariates (51.59\%) is captured by the joint components, these only capture 15.6\% of the variation in the network. This observation suggests that most of the variation in the covariates is associated with the network trading structure, but this information is not enough to capture the variation in the trading patterns.

 \begin{table}
        \centering
		\begin{tabular}{||c c c c||} 
			\hline
			Dataset & Joint & Individual & Residual  \\ [0.5ex] 
			\hline\hline
			Network & 15.60\% & 72.40\% & 12.00\% \\ 
			\hline
			Covariates & 51.59\% & 28.16\% &   20.25\%  \\
			\hline
		\end{tabular}
		\caption{Proportion of variation explained by component for network and covariate datasets}
        \labtable{fig:table:variation_explained}
	\end{table}

	\section{Discussion}
	
	In this work, we have introduced a model for joint and individual information present in network data with node covariates, developed an estimation procedure to obtain joint and individual components of the model, and established theoretical guarantees under a general signal-plus-noise model for the covariates. Illustrations in simulated and real networks show the ability of the methodology to recover meaningful components from the data.
 
 Even though the theory assumes an undirected network with independent Bernoulli entries, similar theoretical guarantees can be obtained when imposing other distributional assumptions on $A$. 
	
	This work is limited to studying only one network and its node covariates. In many situations, more complex data types are also available, such as multiplex or multilayer networks with their corresponding node covariates, or even edge-covariate information. Having access to multiple datasets of this form can yield improvements in the estimation of the joint components.  Extensions of the methodology and theory to these scenarios are left as future work. In this paper, we limit our theoretical analysis to studying the estimation error in Frobenius norm of the joint and individual embeddings. Many applications require further subsequent inferences, including node clustering, prediction or hypothesis testing, for which a more refined analysis of the statistical error and its distributional properties are required, which we leave as future work.

	\section*{Acknowledgement}
 IG was supported by NSF DMS 2422478.

	\appendix
	
	\section{Proofs of main results}
 \subsection{Identifiability}
    \begin{proof} \textbf{[Proof of \reflem{lemma:model:identifiability:1}]}
    We recall that, by definition, $\MM = \MC(P) \cap \MC(W)$, $\MR^{(1)} = \MC(\MP_{\MM^{\perp}} P)$, $\MR^{(2)} = \MC(\MP_{\MM^{\perp}} W)$, $r_M = \dim (\MM)$ and $r_k = \dim (\MR^{(k)})$. Therefore, by construction, $\rnk(P) = r_M + r_1$ and $\rnk(W) = r_M + r_2$. 
    Now, suppose that for $k = 1,2$, $\MM, \MR^{(k)} \neq \{0\}$. 
		\begin{enumerate}[(a)]
			\item \textbf{Existence:} \\
			Define $J^{(1)} = P_{\MM}P$, $J^{(2)} = P_{\MM}W$, $L^{(1)} = \MP_{\MM^{\perp}} P$ and $L^{(2)} = \MP_{\MM^{\perp}} W$. By definition, $\MR^{(k)} = \MC(L^{(k)})$. Then for $k=1,2$,
			\begin{enumerate}
				\item[(1)] $P = J^{(1)} + L^{(1)}$ and $W = J^{(2)} + L^{(2)}$ ,
				\item[(2)] $\MC(J^{(k)}) = \MM$ and $\MM \subset \MC(P) \cap \MC(W)$,
				\item[(3)] $\MM \perp \MC(L^{(k)})$,
				\item[(4)] $\MC(L^{(1)}) \cap \MC(L^{(2)}) = \{0\}$.
			\end{enumerate}
			The proof of Lemma 1 in \citet{Yuan:2022:DMMD} implies that the properties $(1) - (4)$ uniquely identify $J^{(1)}$, $J^{(2)}$, $L^{(1)}$ and $L^{(2)}$.
			Since $\MM \neq \{0\}$, there exists $M \in \O_{n, r_M}$ such that $\MC(M) = \MM$. Since $\MR^{(k)} \neq \{0\}$, there exists $R^{(k)} \in \O_{n, r_k}$ such that $\MC(R^{(k)}) = \MR_k$. We set $\Gam^{(k)}_1 = M^{\top} J^{(k)}$, $\Gam^{(k)}_2 = (R^{(k)})^{\top} L^{(k)}$ and $\Gam^{(k)} = 
			\begin{pmatrix}
				\Gam^{(k)}_1 \\
				\Gam^{(k)}_2
			\end{pmatrix}$.
			Then
			\begin{align*}
				\begin{pmatrix}
					M & R^{(k)}
				\end{pmatrix} \Gam^{(k)} 
				& = 
				\begin{pmatrix}
					M & R^{(k)}
				\end{pmatrix}
				\begin{pmatrix}
					\Gam^{(k)}_1 \\
					\Gam^{(k)}_2
				\end{pmatrix} \\
				& = M \Gam^{(k)}_1 + R^{(k)} \Gam^{(k)}_2 \\
				& = MM^{\top}J^{(k)} + R^{(k)}(R^{(k)})^{\top} L^{(k)} \\
				& = \MP_{\MM} J^{(k)} + \MP_{\MR^{(k)}} L^{(k)} \\
				& = J^{(k)} + L^{(k)} \\
				& = 
                \begin{cases}
                    P, & k = 1, \\
                    W, & k = 2.
                \end{cases}
			\end{align*}
            Since $\begin{pmatrix}
					M & R^{(1)}
				\end{pmatrix} \Gam^{(1)} = P$ and $\MC(P) = \MC \begin{pmatrix}
					M & R^{(1)}
				\end{pmatrix}$,
            we have that 
            \begin{align*}
                \rnk (\Gam^{(1)}) 
                & = \rnk [\begin{pmatrix}
					M & R^{(1)}
				\end{pmatrix}^{\top} P] \\
                & \geq \rnk [\begin{pmatrix}
					M & R^{(1)}
				\end{pmatrix}
                \begin{pmatrix}
					M & R^{(1)}
				\end{pmatrix}^{\top} P] \\
                & = \rnk (\MP_{\MC(P)} P) \\
                & = r_M + r_1.
            \end{align*}
            On the other hand, $\Gam^{(1)} \in \R^{(r_M + r_1) \times n}$. Hence $\rnk(\Gam^{(1)}) \leq r_M + r_1$. Thus $\rnk(\Gam^{(1)}) = r_M + r_1$. Similarly, $\rnk(\Gam^{(2)}) = r_M + r_2$.
            \item \textbf{Orthogonality:} \\
            Since $\MC(R^{(k)}) \subset \MC(M)^{\perp}$, we have that $M \perp R^{(k)}$. 
			\item \textbf{Uniqueness:} \\
			Suppose that there are $N \in \O_{n \times r_M}$, $S^{(k)} \in \O_{n \times r_k}$, $k=1,2$, such that
            \begin{align*}
			& \MC(N) = \MM,
            & \MC(S^{(k)}) = \MR^{(k)}.
		\end{align*}
        Since $\MC(M) = \MM$, there exists full rank $W \in \R^{r_M \times r_M}$ such that $M = NW$. This implies that $N^\top M = W$, and $I_{r_M} = M^\top N W$. Therefore, $W^{-1} = M^\top N = (N^\top M)^\top = W^\top$, that is, the inverse of $W$ is its transpose, which implies that $W$ is an orthogonal matrix. Hence, $M$ is unique up to an orthogonal matrix right multiplication. By similar arguments,  $S^{(k)}=R^{(k)}W'$ for some $W'\in\O_{r_k}$, so $R^{(k)}$ is unique up to right multiplication by an orthogonal matrix.
		\end{enumerate}
	\end{proof}

	\subsection{Spectral method solution}
	\begin{proof} \textbf{[Proof of \refprop{proposition:estimation:spectral:1}]} \labproof{proof:estimation:spectral:1}
		Suppose that $P$ and $W$ satisfy the assumptions in \reflem{lemma:model:identifiability:1} with joint and individual components $(M, R^{(1)}, R^{(2)})$ and ranks $(r_M, r_1, r_2)$. Define $(\widehat{M}, \widehat{R}^{(1)}, \widehat{R}^{(2)})$ to be the output of \refalg{alg:model:spectral:1} when supplied $(P, W, r_M, r_1, r_2)$ as arguments.  \reflem{lemma:model:identifiability:1} implies that there exist $\Gam^{(1)} \in \R^{(r_M + r_1) \times n}$ and $\Gam^{(2)} \in \R^{(r_M + r_2) \times p}$ such that $$P = 
		\begin{pmatrix}
			M & R^{(1)} 
		\end{pmatrix} \Gam^{(1)},
		\quad 
		W = 
		\begin{pmatrix}
			M & R^{(2)}
		\end{pmatrix} \Gam^{(2)}. $$
		Let $\wh{V}^{(1)}, \wh{U}$ be the matrices in the intermediate steps of \refalg{alg:model:spectral:1}. Then, by construction, $\widehat{V}^{(1)} \cong_{O} 
		\begin{pmatrix}
			M & R^{(1)} 
		\end{pmatrix}$ and 
		$\widehat{V}^{(2)} \cong_{O} 
		\begin{pmatrix}
			M & R^{(2)} 
		\end{pmatrix}$. 
		Set $R = 
		\begin{pmatrix}
			R^{(1)} & R^{(2)}
		\end{pmatrix}$. By construction, this $n\times (r_1 + r_2)$ matrix has full rank, and write the eigenvalue decomposition of $RR^{\top}$ as $RR^{\top} = LDL^{\top}$, where $L\in\O_{n,r_1 + r_2}$ and $D\in\mathbb{R}_{(r_1 + r_2)\times (r_1 + r_2)}$ is the diagonal matrix containing the non-zero eigenvalues. Since $M \perp R^{(i)}$, we have that $M \perp L$ and therefore $\widehat{U} \widehat{U}^{\top}$ has eigenvalue decomposition
		\begin{align*}
			\widehat{U} \widehat{U}^{\top}
			& = \widehat{V}^{(1)}(\widehat{V}^{(1)})^{\top} + \widehat{V}^{(2)}(\widehat{V}^{(2)})^{\top} \\
			& = 2MM^{\top} + RR^{\top} \\
			& = 
			\begin{pmatrix}
				M & L
			\end{pmatrix}
			\begin{pmatrix}
				2I_{r_M} & 0 \\
				0 & D
			\end{pmatrix}
			\begin{pmatrix}
				M & L
			\end{pmatrix}^{\top}.
		\end{align*}
		Let $\del = 1 - \sig_1((R^{(1)})^{\top} R^{(2)})$ as in \refeq{eq:estimation:delta}. Denote the leading eigenvalue of $D$ in magnitude by $\lam_1(D)$. We note that we have
        \begin{align*}
			\lam_1(D) 
			& = \lam_1(RR^{\top}) \\
			& = \lam_1(R^{\top}R) \\
			& = \lam_1 \bigg[
			\begin{pmatrix}
				(R^{(1)})^{\top} R^{(1)} & (R^{(1)})^{\top} R^{(2)}  \\
				(R^{(2)})^{\top} R^{(1)} & (R^{(2)})^{\top} R^{(2)}  \\ 
			\end{pmatrix}
			\bigg] \\
			& = \lam_1 \bigg[ 
			\begin{pmatrix}
				I_{r_1} & (R^{(1)})^{\top} R^{(2)}   \\
				(R^{(2)})^{\top} R^{(1)} & I_{r_2}  \\ 
			\end{pmatrix}
			\bigg] \\
            & = \lam_1 \bigg[I_{r_1 + r_2} +  
			\begin{pmatrix}
				0 & (R^{(1)})^{\top} R^{(2)}   \\
				(R^{(2)})^{\top} R^{(1)} & 0  \\ 
			\end{pmatrix}
			\bigg] \\
            & = 1 + \lam_1 \bigg[  
			\begin{pmatrix}
				0 & (R^{(1)})^{\top} R^{(2)}   \\
				(R^{(2)})^{\top} R^{(1)} & 0  \\ 
			\end{pmatrix}
			\bigg] .
		\end{align*}
            Exercise $1.3.17$ in \citet{tao:2023:rand_mat}  implies that
            \begin{align*}
            1 + \lam_1 \bigg[  
			\begin{pmatrix}
				0 & (R^{(1)})^{\top} R^{(2)}   \\
				(R^{(2)})^{\top} R^{(1)} & 0  \\ 
			\end{pmatrix}
			\bigg] 
			& \leq 1 + \sig_1[(R^{(2)})^{\top} R^{(1)}] \\
			& = 1 + \cos(\theta) \\
			& = 2 - \del.
            \end{align*}

        Since $\MC(\MR^{(1)}) \cap \MC(R^{(1)}) = 0$ and $r_M, r_1,r_2 > 0$,
        we have that $\del > 0$. Therefore $\lam_1(D) < 2$ and thus 
		\begin{align*}
			\widehat{M}
			& = \sv(\widehat{U}, r_M) \\
			& \cong_O \eig(\widehat{U} \widehat{U}^{\top}, r_M)  \\
			& \cong_O M.
		\end{align*}
		Then, we have that 
		\begin{align*}
			\widehat{R}^{(1)} 
			& = \sv((I-\widehat{M}\widehat{M}^{\top} )\widehat{V}^{(1)}, r_1 ) \\
			& = \sv(R^{(1)}, r_1) \\
			& \cong_{O} R^{(1)}.
		\end{align*}
		Similarly, $\widehat{R}^{(2)} \cong_{O} R^{(2)}$.
	\end{proof}

	\subsection{Optimization method solution}
	\begin{proof}  \textbf{[Proof of \reflem{lemma:loss_function_minimizer:1}]} 
		Writing $V^{(k)} = 
        \begin{pmatrix}
        M & R^{(k)}
        \end{pmatrix}
        $, the minimization in \eqref{optim:model:main_optim:1} can be rewritten as
		\begin{align}
			\label{eq:model:optimization_fix_indiv:5_sim}
			\minimize_{M \in \O_{n, r_M}}&\{\|A' - \MP_{\MC(V^{(1)})} A'\|_{F}^2 + \|X - \MP_{\MC(V^{(2)})} X\|_{F}^2\}\\
			\mbox{such that} &\quad \quad V^{(k)} = 
			\begin{pmatrix}
				M & R^{(k)}
			\end{pmatrix}  \quad k = 1,2.\nonumber
		\end{align}
		Equivalently, we only need to minimize
		$$
		\|\MP_{\MC(V^{(1)})^{\perp}} A'\|_{F}^2 + \| \MP_{\MC(V^{(2)})^{\perp}} X\|_{F}^2.
		$$
		Due to orthogonality of $R^{(k)}$ and $M$, this is equivalent to minimizing
		$$
		\|\MP_{\MC(R^{(1)})^{\perp}}A' - \MP_{\MC(M)}\MP_{\MC(R^{(1)})^{\perp}}A'\|_{F}^2 + \|\MP_{\MC(R^{(2)})^{\perp}}X - \MP_{\MC(M)}\MP_{\MC(R^{(2)})^{\perp}}X\|_{F}^2.
		$$
		Set
		$$
		Y =
		\begin{pmatrix}
			\MP_{\MC(R^{(1)})^{\perp}} A' & \MP_{\MC(R^{(2)})^{\perp}} X
		\end{pmatrix},
		\quad \quad R = 
		\begin{pmatrix}
			R^{(1)} & R^{(2)}
		\end{pmatrix}.$$
        Then the minimization problem can be equivalently written as
        \begin{align*}
        \minimize_{M \in \O_{n, r_M}} & \left  \{ {\|Y - \MP_{\MC(M)} Y\|_{F}^2} \right \} \\
		\mbox{s.t.} & \quad M^{\top}R = 0.
        \end{align*}
		Because $M$ is orthogonal to $R$, it is equivalent to minimize 
		$$
		\| \MP_{\MC(R)^{\perp}} Y - \MP_{\MC(M)}(\MP_{\MC(R)^{\perp}} Y) \|_{F}^2.
		$$
		The Eckart-Young-Mirsky theorem implies that the optimal $\wh{M}$ is given by $\wh{M} = \sv(\MP_{\MC(R)^{\perp}} Y, r_M)$.
	\end{proof}

	\subsection{Error rate of spectral method}
	\begin{proof} \labproof{proof:main_theorem:1} \textbf{[Proof of \refthm{theorem:error_rate:1}]}\\  
		We split the proof intro three parts, covering the error rate for each of the joint, network individual and covariate individual components. 
		\begin{itemize}
			\item \tbf{Rate for Joint Components:} \\
            We recall from Definition 5.13 in \citet{nonasymp_analysis}, the $\psi_1$-norm of a random variable $T$ is given by $\| T\|_{\psi_1} = \sup\limits_{q \geq 1} [q^{-1} (\E|T|^q)^{1/q}]$. We recall that
			\begin{align*}
				& \ep^{(1)} = \frac{\sqrt{r_M+r_1} \kappa}{\lam_{r_M + r_1}(P)},
				& \ep^{(2)} = \frac{\tau \sqrt{n (\sig_{r_M + r_2}^2(W) + p ) (r_M + r_2)}}{\sig_{r_M + r_2}^2(W)} \wedge \sqrt{r_M + r_2}, \\
				& \del = 1-\cos(\theta),
			\end{align*}
            where $r_M = \dim(\MM)$, $r_k = \dim(\MR^{(k)})$ and $\theta$ is the first principal angle between $\MC(R^{(1)})$ and $\MC(R^{(2)})$ so that $\cos(\theta) = \sig_1((R^{(2)})^{\top} R^{(1)})$. According to \reflem{lemma:model:identifiability:1}, there exist $M \in \R^{n \times r_M}$, $R^{(1)} \in \R^{n \times r_1}$, $R^{(2)} \in \R^{n \times r_2}$, $\Gam^{(1)} \in \R^{(r_M + r_1) \times n}$ and  $\Gam^{(2)} \in \R^{(r_M + r_2) \times p}$ such that 
            \begin{align}
            &P = 
				\begin{pmatrix}
					M & R^{(1)} 
				\end{pmatrix} \Gam^{(1)},
			& W = 
				\begin{pmatrix}
					M & R^{(2)} 
				\end{pmatrix} \Gam^{(2)},  \labeq{proof:thm1:factorization}
            \end{align}
		    $M \perp R^{(k)}$ and $M$, $\MR^{(1)}$, $R^{(2)}$ are unique modulo $\cong_{O}$.  
			 For notational convenience, we set $A^{(1)} = A$, $A^{(2)} = X$ and $P^{(k)} = \E(A^{(k)})$. Following \refalg{alg:model:spectral:1} we define 
			\begin{align*}
				& \widehat{V}^{(k)}  = \sv(A^{(k)}, r_M + r_i),
				& V^{(k)} = \sv(P^{(k)}, r_M + r_i), \\
				& \widehat{U} = 
				\begin{pmatrix}
					\widehat{V}^{(1)} & \widehat{V}^{(2)}
				\end{pmatrix},
				& U = 
				\begin{pmatrix}
					V^{(1)} & V^{(2)}
				\end{pmatrix}, \\
				& \widehat{M} = \sv(\widehat{U}, r_M),
				& \widehat{R}^{(k)} = \sv(\MP_{\widehat{\MM}^{\perp}} \widehat{V}^{(k)}, r_k).
			\end{align*}
            From the proof of \refprop{proposition:estimation:spectral:1}, we know that
            $V^{(k)} \cong_{O} \begin{pmatrix}
            M & R^{(k)}
            \end{pmatrix}$, $M \cong_{O} \sv(U, r_M)$ and $R^{(k)} \cong_{O} \sv(\MP_{\MM^{\perp}} V^{(k)}, r_k)$.
            Since we need only consider the projection matrix onto a components respective column space and such a projection matrix is invariant under right multiplication by an orthogonal matrix, we may assume that $V^{(k)} = \begin{pmatrix}
            M & R^{(k)}
            \end{pmatrix}$, $M = \sv(U, r_M)$ and $R^{(k)} = \sv(\MP_{\MM^{\perp}} V^{(k)}, r_k)$ so that
            $P = V^{(1)} \Gam^{(1)}$ and $W = V^{(2)} \Gam^{(2)}$.
            We will work with corresponding projection matrices 
			\begin{align*}
				& \MP_{\widehat{\MV}^{(k)}} =  \widehat{V}^{(k)}(\widehat{V}^{(k)})^{\top},
				& \MP_{\MV^{(k)}} =  V^{(k)}(V^{(k)})^{\top}, \\ 
				& \MP_{\widehat{\MM}} = \widehat{M} \widehat{M}^{\top},
				& \MP_{\MM} = MM^{\top}, \\
				& \MP_{\widehat{\MR}^{(k)}} = \widehat{R}^{(k)} (\widehat{R}^{(k)})^{\top},
				& \MP_{\MR^{(k)}} = R^{(k)}(R^{(k)})^{\top},
			\end{align*}
            and we define
            $$\MQ_{\widehat{\MU}} = \widehat{U} \widehat{U}^{\top},
				\quad \quad \MQ_{\MU} = UU^{\top}.$$
			With those definitions in mind, we now begin deriving our error bound. From \refeq{proof:thm1:factorization}, we know that $V^{(k)} \cong_{O} 
			\begin{pmatrix}
				M & R^{(k)} 
			\end{pmatrix}$. By defintion, 
			\begin{align*}
				\MQ_{\widehat{\MU}} 
				& = \widehat{U} \widehat{U}^{\top} \\
				& =  
				\begin{pmatrix}
					\widehat{V}^{(1)} & \widehat{V}^{(2)}
				\end{pmatrix}
				\begin{pmatrix}
					(\widehat{V}^{(1)})^{\top} \\ 
					(\widehat{V}^{(2)})^{\top}
				\end{pmatrix} \\
				& = \MP_{\widehat{\MV}^{(1)}} + \MP_{\widehat{\MV}^{(2)}}.
			\end{align*}
			Similarly, $\MQ_{\MU} = \MP_{\MV^{(1)}} + \MP_{\MV^{(2)}}$.
			Finally, we set 
			\[
			R = 
			\begin{pmatrix}
				R^{(1)} & R^{(2)}
			\end{pmatrix}
			\]
			and we write the eigendecomposition of 
			$RR^{\top}$ as $$RR^{\top} = L D L^{\top}.$$ 
            As in the proof of \refprop{proposition:estimation:spectral:1}, we have that the eigenvalue decomposition of $\MQ_{\MU}$ is given by
			\begin{align*}
				\MQ_{\MU} 
				&= \MP_{\MV^{(1)}} + \MP_{\MV^{(2)}} \\
				&= 2MM^{\top} + LDL^{\top} \\
				&= 
				\begin{pmatrix}
					M & L
				\end{pmatrix}
				\begin{pmatrix}
					2I & 0  \\
					0 & D  \\ 
				\end{pmatrix}
				\begin{pmatrix}
					M & L
				\end{pmatrix}^{\top}
			\end{align*}
			so that
		\begin{align*}
			\lam_1(D) 
            & = 1 + \lam_1 \bigg[  
			\begin{pmatrix}
				0 & (R^{(1)})^{\top} R^{(2)}   \\
				(R^{(2)})^{\top} R^{(1)} & 0  \\ 
			\end{pmatrix}
			\bigg]
		\end{align*}
            and
            \begin{align*}
            1 + \lam_1 \bigg[  
			\begin{pmatrix}
				0 & (R^{(1)})^{\top} R^{(2)}   \\
				(R^{(2)})^{\top} R^{(1)} & 0  \\ 
			\end{pmatrix}
			\bigg] 
			& = 2 - \del.
            \end{align*}
			Since $\del > 0$, we see that $\lam_1(D) < 2$ and thus $\eig(\MQ_{\MU}, r_M) = M$. In addition,
			\begin{align*}
				\lam_{r_M}(\MQ_{\MU}) - \lam_{r_M + 1}(\MQ_{\MU}) 
				& = 2 - \lam_1(D) \\
				& = 2 - (2 - \del) \\
				& = \del.
			\end{align*}
			Therefore
            $$\frac{1}{\lam_{r_M}(\MQ_{\MU}) - \lam_{r_M + 1}(\MQ_{\MU}) } \leq \frac{1}{\del}.$$
            From here, Lemma $2.6$ in \citet{Chen:2021:spec_methods_data_sci} and Theorem $2$ in \citet{Yu:2015:davis_kahan_for_stats} imply that that for some $b_0 > 0$,
			\begin{align*}
				\E[ \min\limits_{Q \in \O_{r_M}} \|\widehat{M} - MQ\|_F ]
				&\leq \frac{b_0\sqrt{r_M}\E\| \MQ_{\widehat{\MU}} - \MQ_{\MU}\|_F}{\lam_{r_M}(\MQ_{\MU}) - \lam_{r_M + 1}(\MQ_{\MU})} \\
                & \leq \frac{b_0 \sqrt{r_M}}{\del}\E\| \MQ_{\widehat{\MU}} - \MQ_{\MU}\|_F.
			\end{align*}
            To bound $\|\MQ_{\widehat{\MU}} - \MQ_{\MU}\|_F$, we note that  
			\begin{align*}
				\|\MQ_{\widehat{\MU}} - \MQ_{\MU}\|_F
				&= \bigg \|\sum_{i=1}^2 \MP_{\widehat{\MV}^{(i)}} - \MP_{\MV^{(i)}}\bigg \|_F \\
				& \leq \sum_{i=1}^2 \|\MP_{\widehat{\MV}^{(i)}} - \MP_{\MV^{(i)}}\|_F. \\
			\end{align*}
   
			We make some useful observations about the singular vectors of $\E(XX^{\top})$ and $W$.
			First, by definition, $\eig(P, r_M + r_1) \cong_{O} V^{(1)}$.  Second, from \refeq{proof:thm1:factorization}, we know that    
			\begin{align*}
				\E(XX^{\top})  
				&= \E((X-W)(X - W)^{\top}) + W W^{\top} \\
				&= p\tau^2I_n + WW^{\top}.
			\end{align*}

			Let $\Gam^{(2)} (\Gam^{(2)})^{\top}$ have eigendecomposition $\Gam^{(2)} (\Gam^{(2)})^{\top} = Q \Lam Q^{\top}$. Then we write 
			\begin{align*}
				WW^{\top} 
				&= V^{(2)}\Gam^{(2)} (V^{(2)} \Gam^{(2)})^{\top} \\
				&= V^{(2)}Q \Lam (V^{(2)}Q)^{\top} \\
				&= \begin{pmatrix}
					V^{(2)} & V^{(2)}_{\perp}
				\end{pmatrix}
				\begin{pmatrix}
					Q & 0 \\
					0 & I_{n - r_M - r_2}
				\end{pmatrix}
				\begin{pmatrix}
					\Lam & 0 \\
					0 & 0
				\end{pmatrix}
				\bigg[ \begin{pmatrix}
					V^{(2)} & V^{(2)}_{\perp}
				\end{pmatrix}
				\begin{pmatrix}
					Q & 0 \\
					0 & I_{n - r_M - r_2}
				\end{pmatrix} \bigg]^{\top}.
			\end{align*}
			Since $\eig(\E(XX^{\top}), r_m + r_2) = \eig(WW^{\top}, r_m + r_2)$ with corresponding eigenvalues $\lam_i(\E(XX^{\top})) = p\tau^2 + \lam_i(WW^{\top})$, we have that  
			\begin{align*}
				\eig(\E(XX^{\top}), r_M + r_2) 
				&= \eig(WW^{\top}, r_m + r_2) \\
				&= \eig \bigg( 
				\begin{pmatrix}
					V^{(2)} & V^{(2)}_{\perp}
				\end{pmatrix}
				\begin{pmatrix}
					Q & 0 \\
					0 & I_{n - r_M - r_i}
				\end{pmatrix}
				\bigg[ \begin{pmatrix}
					V^{(2)} & V^{(2)}_{\perp}
				\end{pmatrix}
				\begin{pmatrix}
					Q & 0 \\
					0 & I_{n - r_M - r_2}
				\end{pmatrix} \bigg]^{\top}, 
				r_M + r_2
				\bigg) \\
				&= \eig (V^{(2)}QQ^{\top}(V^{(2)})^{\top} + V^{(2)}_{\perp} (V^{(2)}_{\perp})^{\top}, r_M + r_2) \\
				&= \eig (\MP_{\MV^{(2)}} + V^{(2)}_{\perp} (V^{(2)}_{\perp})^{\top}, r_M + r_2) \\
				&= \eig \bigg(
				\begin{pmatrix}
					V^{(2)} & V^{(2)}_{\perp}
				\end{pmatrix}
				\begin{pmatrix}
					V^{(2)} & V^{(2)}_{\perp}
				\end{pmatrix}^{\top} , r_M + r_2 \bigg) \\
				& \cong_{O} V^{(2)}.
			\end{align*}

        Since $\eig(P, r_M + r_1) = V^{(1)}$, we apply the Davis-Kahan theorem to obtain that for some $b_1 > 0$
			\begin{align*}
				\|\MP_{\widehat{\MV}^{(1)}} - \MP_{\MV^{(1)}}\|_F 
				&\leq \frac{b_1 \sqrt{r_M + r_1} \|A - P \|}{\lam_{r_M + r_1}(P)}. 
			\end{align*} 
        Since $\|A - P\|$ is subexponential and for each $q \geq 1$, 
			$$\E(\| A - P\|^q)^{1/q} \leq \bigg\| \| A - P\| \bigg \|_{\psi_1} q. $$ 
			
			By \refasmp{asmp:estimation:distribution_general}, for each $q \geq 1$, $\bigg \| \|A - P\| \bigg \|_{\psi_1} \leq \kappa$. Therefore
			\begin{align*}
				\mathbb{E} \bigg( \|\MP_{\widehat{\MV}^{(1)}} - \MP_{\MV^{(1)}} \|_F^q \bigg)^{1/q}
				& \leq \E \bigg[ \frac{b_1^q (r_M + r_1)^{q/2} \| A - P\|^q}{(\lam_{r_M + r_1}(P))^q} \bigg] ^{1/q} \\
				&= \frac{b_1 \sqrt{r_M + r_1} \E(\| A - P\|^q)^{1/q}}{\lam_{r_M + r_1}(P)} \\
				&\leq \frac{b_1 \sqrt{r_M + r_1} \bigg\| \|A - P \| \bigg\|_{\psi_1}  q}{\lam_{r_M + r_1}(P)} \\
                &\leq \frac{b_1 \sqrt{r_M + r_1} \kappa q}{\lam_{r_M + r_1}(P)}.
			\end{align*}
			This implies that
			\begin{align*}
				\E \|\MP_{\widehat{\MV}^{(1)}} - \MP_{\MV^{(1)}}\|_F 
				& \leq  \bigg\| \|\MP_{\widehat{\MV}^{(1)}} - \MP_{\MV^{(1)}}\|_F \bigg\|_{\psi_1} \\
				&= \sup_{q \geq 1} \bigg[ \frac{\E(\|\MP_{\widehat{\MV}^{(1)}} - \MP_{\MV^{(1)}}\|_F^q)^{1/q}}{q} \bigg] \\
				&\leq \frac{b_1 \sqrt{r_M + r_1} \kappa }{\lam_{r_M + r_1}(P)} \\
				& = O\bigg( \ep^{(1)} \bigg).
			\end{align*}
			Jensen's inequality implies that $\E\|\MP_{\widehat{\MV}^{(2)}} - \MP_{\MV^{(2)}}\|_F \leq \bigg(\E[\|\MP_{\widehat{\MV}^{(2)}} - \MP_{\MV^{(2)}}\|_F^2] \bigg)^{1/2}$. Under \refasmp{asmp:estimation:distribution_general}, Theorem 3 in \citet{Cai:2018:rate_optim_perturb_bound_sing_subspace} implies that there exists $c_0 > 0$ such that
			\begin{align*}
				\E[\|\MP_{\widehat{\MV}^{(2)}} - \MP_{\MV^{(2)}}\|_F^2]
                & \leq \frac{c_0 n (\sig_{r_M + r_2}^2(\tau^{-1} W) + p ) (r_M + r_2)}{\sig_{r_M + r_2}^4(\tau^{-1}W)} \wedge (r_M + r_2)  \\
				& = \frac{c_0 \tau^2 n (\sig_{r_M + r_2}^2(W) + \tau^2p ) (r_M + r_2)}{\sig_{r_M + r_2}^4(W)} \wedge (r_M + r_2) .
			\end{align*}
			Therefore, 
			\begin{align*}
				\E\|\MP_{\widehat{\MV}^{(2)}} - \MP_{\MV^{(2)}}\|_F 
				& \leq \frac{\tau \sqrt{c_0 n (\sig_{r_M + r_2}^2(W) + \tau^2 p ) (r_M + r_2)}}{\sig_{r_M + r_2}^2(W)} \wedge \sqrt{r_M + r_2} \\
				& =  O\bigg( \ep^{(2)} \bigg).
			\end{align*}
			We therefore obtain
			\begin{align*}
				\E \|\MQ_{\widehat{\MU}} - \MQ_{\MU}\|_F 
				&=  \E\bigg \| \sum_{i=1}^2 \MP_{\widehat{\MV}^{(i)}} - \MP_{\MV^{(i)}} \bigg\|_F \\
				& \leq  \sum_{i=1}^2 \E\|\MP_{\widehat{\MV}^{(i)}} - \MP_{\MV^{(i)}}\|_F \\
				& = O\bigg( \ep^{(1)} + \ep^{(2)} \bigg). 
			\end{align*}
			Summarizing, we have that
			\begin{align*}
				 \E[ d(\wh{M}, M)]
                & = \E[ \min\limits_{Q \in \O_{r_M}} \|\widehat{M} - MQ\|_F ] \\
                & = O\bigg(  \frac{\sqrt{r_M}}{\del}\E\| \MQ_{\widehat{\MU}} - \MQ_{\MU}\|_F \bigg) \\
				& = O\bigg(  \frac{\sqrt{r_M}}{\del}   [ \ep^{(1)} +  \ep^{(2)}] \bigg).
			\end{align*}
			\item \tbf{Rate for Network Individual Components:} \\
                We note that
			\begin{align*}
				\MP_{\widehat{\MM}^{\perp}} \MP_{\widehat{\MV}^{(1)}} \MP_{\widehat{\MM}^{\perp}} - \MP_{\MM^{\perp}} \MP_{\MV^{(1)}} \MP_{\MM^{\perp}}
				& = \bigg( \MP_{\widehat{\MM}^{\perp}} \MP_{\widehat{\MV}^{(1)}} \MP_{\widehat{\MM}^{\perp}} - \MP_{\MM^{\perp}} \MP_{\widehat{\MV}^{(1)}} \MP_{\widehat{\MM}^{\perp}} \bigg) \\
				& \hspace{1cm}+ \bigg( \MP_{\MM^{\perp}} \MP_{\widehat{\MV}^{(1)}} \MP_{\widehat{\MM}^{\perp}} - \MP_{\MM^{\perp}} \MP_{\MV^{(1)}} \MP_{\MM^{\perp}} \bigg) \\
				& =  (\MP_{\widehat{\MM}^{\perp}} - \MP_{\MM^{\perp}} )  \MP_{\widehat{\MV}^{(1)}} \MP_{\widehat{\MM}^{\perp}} \\ 
				& \hspace{1cm} + \MP_{\MM^{\perp}}(\MP_{\widehat{\MV}^{(1)}} \MP_{\widehat{\MM}^{\perp}} -  \MP_{\MV^{(1)}} \MP_{\MM^{\perp}}).
			\end{align*}
			The expression in parentheses in the second term can be further expanded as
			\begin{align*}
				\MP_{\widehat{\MV}^{(1)}} \MP_{\widehat{\MM}^{\perp}} -  \MP_{\MV^{(1)}} \MP_{\MM^{\perp}}
				& = \bigg( \MP_{\widehat{\MV}^{(1)}} \MP_{\widehat{\MM}^{\perp}} - \MP_{\widehat{\MV}^{(1)}} \MP_{\MM^{\perp}} \bigg) + \bigg(\MP_{\widehat{\MV}^{(1)}} \MP_{\MM^{\perp}} - \MP_{\MV^{(1)}} \MP_{\MM^{\perp}} \bigg) \\
				& = \MP_{\widehat{\MV}^{(1)}} (\MP_{\widehat{\MM}^{\perp}} - \MP_{\MM^{\perp}} ) + (\MP_{\widehat{\MV}^{(1)}} - \MP_{\MV^{(1)}})\MP_{\MM^{\perp}}. 
			\end{align*}
			Since the spectral norm of a projection matrix is equal to one, we have that
			\begin{align*}
				\|\MP_{\widehat{\MM}^{\perp}} \MP_{\widehat{\MV}^{(1)}} \MP_{\widehat{\MM}^{\perp}} - \MP_{\MM^{\perp}} \MP_{\MV^{(1)}} \MP_{\MM^{\perp}}\|
				& \leq \|\MP_{\widehat{\MM}^{\perp}} - \MP_{\MM^{\perp}}\| \|\MP_{\widehat{\MV}^{(1)}}\|  \|\MP_{\widehat{\MM}^{\perp}}\| \\
				& \hspace{1cm} + \|\MP_{\MM^{\perp}}\|\|\MP_{\widehat{\MV}^{(1)}}\| \| \MP_{\widehat{\MM}^{\perp}} - \MP_{\MM^{\perp}}\| \\
				& \hspace{1cm} + \|\MP_{\MM^{\perp}}\| \|\MP_{\widehat{\MV}^{(1)}} - \MP_{\MV^{(1)}}\| \|\MP_{\MM^{\perp}}\|  \\
				& \leq 2\|\MP_{\widehat{\MM}^{\perp}} - \MP_{\MM^{\perp}}\|  + \|\MP_{\widehat{\MV}^{(1)}} - \MP_{\MV^{(1)}}\|  \\
				& = 2 \|\MP_{\widehat{\MM}} - \MP_{\MM}\| + \|\MP_{\widehat{\MV}^{(1)}} - \MP_{\MV^{(1)}}\| .
			\end{align*}
			Since 
			\begin{align*}
				\widehat{R}^{(1)} 
				& = \sv(\MP_{\widehat{\MM}^{\perp}} \widehat{V}^{(1)}, r_1) \\
				& = \eig(\MP_{\widehat{\MM}^{\perp}} \widehat{V}^{(1)} 
				(\MP_{\widehat{\MM}^{\perp}} \widehat{V}^{(1)})^{\top}, r_1) \\
				& = \eig(\MP_{\widehat{\MM}^{\perp}} \MP_{\widehat{\MV}^{(1)}} \MP_{\widehat{\MM}^{\perp}}, r_1 ),
			\end{align*}
			\begin{align*}
				R^{(1)} 
				& = \sv(\MP_{\MM^{\perp}} V^{(1)}, r_1) \\
				& = \eig(\MP_{\MM^{\perp}} V^{(1)}(\MP_{\MM^{\perp}} V^{(1)})^{\top}, r_1) \\
				& = \eig(\MP_{\MM^{\perp}} \MP_{\MV^{(1)}} \MP_{\MM^{\perp}}, r_1 )
			\end{align*}
			and 
			\begin{align*}
				\MP_{\MM^{\perp}} \MP_{\MV^{(1)}} \MP_{\MM^{\perp}}
				& = \MP_{\MM^{\perp}} (MM^{\top} + R^{(1)}(R^{(1)})^{\top} ) \MP_{\MM^{\perp}} \\
				& = R^{(1)}(R^{(1)})^{\top} \\
				& = \MP_{\MR^{(1)}},
			\end{align*}
			the Davis-Kahan theorem and the fact that orthogonal matrices have unit nontrivial singular values imply that 
			\begin{align*}
				\| \widehat{R^{(1)}} (\widehat{R^{(1)}} )^{\top} - R^{(1)} (R^{(1)})^{\top} \|_F  
				& \leq 2 \sqrt{r_1} \frac{\|\MP_{\widehat{\MM}^{\perp}} \MP_{\widehat{\MV}^{(1)}} \MP_{\widehat{\MM}^{\perp}} - \MP_{\MM^{\perp}} \MP_{\MV^{(1)}} \MP_{\MM^{\perp}}\|}{\lam_{r_1}(\MP_{R^{(1)}})} \\
                & = 2 \sqrt{r_1} \frac{\|\MP_{\widehat{\MM}^{\perp}} \MP_{\widehat{\MV}^{(1)}} \MP_{\widehat{\MM}^{\perp}} - \MP_{\MM^{\perp}} \MP_{\MV^{(1)}} \MP_{\MM^{\perp}}\|}{\sig_{r_1}(R^{(1)})^2}  \\
                & = 2 \sqrt{r_1}\|\MP_{\widehat{\MM}^{\perp}} \MP_{\widehat{\MV}^{(1)}} \MP_{\widehat{\MM}^{\perp}} - \MP_{\MM^{\perp}} \MP_{\MV^{(1)}} \MP_{\MM^{\perp}}\|
			\end{align*}
			and that 
			\begin{align*}
				\E\|\MP_{\widehat{\MV}^{(1)}} - \MP_{\MV^{(1)}}\| 
				& \leq \frac{\sqrt{2} \E \|A - P\|}{\lam_{r_M + r_1}(P)} \\
				& \leq \frac{ \sqrt{2} \bigg \|\|A - P\| \bigg \|_{\psi_1}}{\lam_{r_M + r_1}(P)} \\
                & \leq \frac{ \sqrt{2} \sqrt{r_M + r_1} \kappa}{\lam_{r_M + r_1}(P)} \\
				& = \sqrt{2} \ep^{(1)}.
			\end{align*}
			Hence 
			\begin{align*}
				\E [d(\wh{R}^{(1)}, R^{(1)})] 
                & = \E \bigg[\min_{Q \in \O_{r_1}} \|R^{(1)} - \widehat{R}^{(1)}Q\|_F \bigg] \\
				& = O\bigg( \sqrt{r_1} \E \| \MP_{\widehat{\MM}^{\perp}} \MP_{\widehat{\MV}^{(1)}} \MP_{\widehat{\MM}^{\perp}} - \MP_{\MM^{\perp}} \MP_{\MV^{(1)}} \MP_{\MM^{\perp}}\| \bigg) \\
				& =  O\bigg(  \sqrt{r_1} \bigg[  \E \|\MP_{\widehat{\MM}} - \MP_{\MM}\| + \E \|\MP_{\widehat{\MV}^{(1)}} - \MP_{\MV^{(1)}}\|  \bigg] \bigg) \\
				& = O\bigg(  \sqrt{r_M r_1} \bigg[ \frac{\ep^{(1)} +  \ep^{(2)}}{\del}   + \ep^{(1)} \bigg]  \bigg) \\
                & = O\bigg( \sqrt{r_M r_1} \bigg[ \frac{\ep^{(1)} +  \ep^{(2)}}{\del} \bigg]  \bigg) \\
				& = O\bigg(  \frac{\sqrt{r_M r_1}}{\del } \bigg[ \ep^{(1)} +  \ep^{(2)} \bigg]  \bigg). \\
			\end{align*}
			\item \tbf{Rate for Covariate Individual Components:} \\
            Similarly to the network individual components, we note that
			\begin{align*}
				\MP_{\widehat{\MM}^{\perp}} \MP_{\widehat{\MV}^{(2)}} \MP_{\widehat{\MM}^{\perp}} - \MP_{\MM^{\perp}} \MP_{\MV^{(2)}} \MP_{\MM^{\perp}}
				& = \bigg( \MP_{\widehat{\MM}^{\perp}} \MP_{\widehat{\MV}^{(2)}} \MP_{\widehat{\MM}^{\perp}} - \MP_{\MM^{\perp}} \MP_{\widehat{\MV}^{(2)}} \MP_{\widehat{\MM}^{\perp}} \bigg) \\
				& \hspace{1cm}+ \bigg( \MP_{\MM^{\perp}} \MP_{\widehat{\MV}^{(2)}} \MP_{\widehat{\MM}^{\perp}} - \MP_{\MM^{\perp}} \MP_{\MV^{(2)}} \MP_{\MM^{\perp}} \bigg) \\
				& =  (\MP_{\widehat{\MM}^{\perp}} - \MP_{\MM^{\perp}} )  \MP_{\widehat{\MV}^{(2)}} \MP_{\widehat{\MM}^{\perp}} \\ 
				& \hspace{1cm} + \MP_{\MM^{\perp}}(\MP_{\widehat{\MV}^{(2)}} \MP_{\widehat{\MM}^{\perp}} -  \MP_{\MV^{(2)}} \MP_{\MM^{\perp}}).
			\end{align*}
			The expression in parentheses in the second term can be further expanded as
			\begin{align*}
				\MP_{\widehat{\MV}^{(2)}} \MP_{\widehat{\MM}^{\perp}} -  \MP_{\MV^{(2)}} \MP_{\MM^{\perp}}
				& = \bigg( \MP_{\widehat{\MV}^{(2)}} \MP_{\widehat{\MM}^{\perp}} - \MP_{\widehat{\MV}^{(2)}} \MP_{\MM^{\perp}} \bigg) + \bigg(\MP_{\widehat{\MV}^{(2)}} \MP_{\MM^{\perp}} - \MP_{\MV^{(2)}} \MP_{\MM^{\perp}} \bigg) \\
				& = \MP_{\widehat{\MV}^{(2)}} (\MP_{\widehat{\MM}^{\perp}} - \MP_{\MM^{\perp}} ) + (\MP_{\widehat{\MV}^{(2)}} - \MP_{\MV^{(2)}})\MP_{\MM^{\perp}}. 
			\end{align*}
			Since the spectral norm of a projection matrix is $1$, we have that
			\begin{align*}
				\|\MP_{\widehat{\MM}^{\perp}} \MP_{\widehat{\MV}^{(2)}} \MP_{\widehat{\MM}^{\perp}} - \MP_{\MM^{\perp}} \MP_{\MV^{(2)}} \MP_{\MM^{\perp}}\|
				& \leq \|\MP_{\widehat{\MM}^{\perp}} - \MP_{\MM^{\perp}}\| \|\MP_{\widehat{\MV}^{(2)}}\|  \|\MP_{\widehat{\MM}^{\perp}}\| \\
				& \hspace{1cm} + \|\MP_{\MM^{\perp}}\|\|\MP_{\widehat{\MV}^{(2)}}\| \| \MP_{\widehat{\MM}^{\perp}} - \MP_{\MM^{\perp}}\| \\
				& \hspace{1cm} + \|\MP_{\MM^{\perp}}\| \|\MP_{\widehat{\MV}^{(2)}} - \MP_{\MV^{(2)}}\| \|\MP_{\MM^{\perp}}\| \\
				& \leq 2\|\MP_{\widehat{\MM}^{\perp}} - \MP_{\MM^{\perp}}\|  + \|\MP_{\widehat{\MV}^{(2)}} - \MP_{\MV^{(2)}}\| \\
				& = 2 \|\MP_{\widehat{\MM}} - \MP_{\MM}\| + \|\MP_{\widehat{\MV}^{(2)}} - \MP_{\MV^{(2)}}\| .
			\end{align*}
			Since 
			\begin{align*}
				\widehat{R}^{(2)} 
				& = \sv(\MP_{\widehat{\MM}^{\perp}} \widehat{V}^{(2)}, r_2) \\
				& = \eig(\MP_{\widehat{\MM}^{\perp}} \widehat{V}^{(2)} 
				(\MP_{\widehat{\MM}^{\perp}} \widehat{V}^{(2)})^{\top}, r_2) \\
				& = \eig(\MP_{\widehat{\MM}^{\perp}} \MP_{\widehat{\MV}^{(2)}} \MP_{\widehat{\MM}^{\perp}}, r_2 ),
			\end{align*}
			\begin{align*}
				R^{(2)} 
				& = \sv(\MP_{\MM^{\perp}} V^{(2)}, r_2) \\
				& = \eig(\MP_{\MM^{\perp}} V^{(2)}(\MP_{\MM^{\perp}} V^{(2)})^{\top}, r_2) \\
				& = \eig(\MP_{\MM^{\perp}} \MP_{\MV^{(2)}} \MP_{\MM^{\perp}}, r_2 )
			\end{align*}
			and 
			\begin{align*}
				\MP_{\MM^{\perp}} \MP_{\MV^{(2)}} \MP_{\MM^{\perp}}
				& = \MP_{\MM^{\perp}} (MM^{\top} + R^{(2)}(R^{(2)})^{\top} ) \MP_{\MM^{\perp}} \\
				& = R^{(2)}(R^{(2)})^{\top} \\
				& = \MP_{\MR^{(2)}},
			\end{align*}
			the Davis-Kahan theorem and the fact that orthogonal matrices have unit nontrivial singular values imply that
			\begin{align*}
				\| \widehat{R^{(2)}} (\widehat{R^{(2)}} )^{\top} - R^{(2)} (R^{(2)})^{\top} \|_F  
				& \leq 2 \sqrt{r_2} \frac{\|\MP_{\widehat{\MM}^{\perp}} \MP_{\widehat{\MV}^{(2)}} \MP_{\widehat{\MM}^{\perp}} - \MP_{\MM^{\perp}} \MP_{\MV^{(2)}} \MP_{\MM^{\perp}}\|}{\lam_{r_2}(\MP_{\MR^{(2)}})} \\
				& = 2 \sqrt{r_2} \frac{\|\MP_{\widehat{\MM}^{\perp}} \MP_{\widehat{\MV}^{(2)}} \MP_{\widehat{\MM}^{\perp}} - \MP_{\MM^{\perp}} \MP_{\MV^{(2)}} \MP_{\MM^{\perp}}\|}{\sig_{r_2}(R^{(2)})^2} \\
                & = 2 \sqrt{r_2} \|\MP_{\widehat{\MM}^{\perp}} \MP_{\widehat{\MV}^{(2)}} \MP_{\widehat{\MM}^{\perp}} - \MP_{\MM^{\perp}} \MP_{\MV^{(2)}} \MP_{\MM^{\perp}}\|.
			\end{align*}
			Theorem 3 in \citet{Cai:2018:rate_optim_perturb_bound_sing_subspace} then implies that there exists $c_0 >0$ such that
			\begin{align*}
				\E \|\MP_{\widehat{\MV}^{(2)}} - \MP_{\MV^{(2)}}\| 
				& \leq (\E\|\MP_{\widehat{\MV}^{(2)}} - \MP_{\MV^{(2)}}\|^2)^{1/2} \\
				& \leq \bigg[ \frac{c_0 \tau^2 n (\sig_{r_M + r_2}(W)^2 + p)}{\sig_{r_M + r_2}(W)^{4}} \wedge 1 \bigg]^{1/2}  \\
                & \leq \sqrt{r_M+r_2} \bigg[ \frac{c_0 \tau^2 n (\sig_{r_M + r_2}(W)^2 + p)}{\sig_{r_M + r_2}(W)^{4}} \wedge 1 \bigg]^{1/2}  \\
				& = \sqrt{c_0} \ep^{(2)} .
			\end{align*}
			Hence 
			\begin{align*}
				\E [d(\wh{R}^{(1)}, R^{(2)})] 
                & = \E \bigg[\min_{Q \in \O_{r_2}} \|R^{(2)} - \widehat{R}^{(2)}Q\|_F \bigg] \\
				& = O\bigg( \sqrt{r_2} \E \| \MP_{\widehat{\MM}^{\perp}} \MP_{\widehat{\MV}^{(2)}} \MP_{\widehat{\MM}^{\perp}} - \MP_{\MM^{\perp}} \MP_{\MV^{(2)}} \MP_{\MM^{\perp}}\| \bigg) \\
				& =  O\bigg( \sqrt{r_2} \bigg[  \E \|\MP_{\widehat{\MM}} - \MP_{\MM}\| + \E \|\MP_{\widehat{\MV}^{(2)}} - \MP_{\MV^{(2)}}\|  \bigg] \bigg) \\
				& = O\bigg(  \sqrt{r_M r_2}  \bigg[ \frac{\ep^{(1)} +  \ep^{(2)}}{\del}   + \ep^{(2)} \bigg]  \bigg) \\
                & = O\bigg(  \sqrt{r_M r_2}  \bigg[ \frac{\ep^{(1)} +  \ep^{(2)}}{\del} \bigg]  \bigg) \\
				& = O\bigg(  \frac{\sqrt{r_M r_2}}{\del} \bigg[ \ep^{(1)} +  \ep^{(2)} \bigg]  \bigg). \\
			\end{align*}
		\end{itemize}
		
	\end{proof}

    \subsection{Proof of \refcor{corollary:Estimation Error:consistency:2}}
    \begin{proof} \textbf{[Proof of \refcor{corollary:Estimation Error:consistency:2}]} 
    From \citet{nonasymp_analysis}, we know that $\bigg \| \|A-P\| \bigg\|_{\psi_1}$ differs from $\kappa$ by a multiplicative constant. Since $\mu(P) = \Om(\rho_n n)$ and $\rho_n = \om(n^{-1} \log n)$, we have that $\mu(P) = \om(\log(n))$. The proof of Lemma 13 in \citet{Arroyo:2021:cosie} implies that 
    \begin{align*}
    \bigg \| \|P - A\|\bigg \|_{\psi_1} 
    & = O \bigg( \sqrt{\mu(P)} \bigg)  \\
    & = O\bigg( \sqrt{\rho_n n} \bigg).  
    \end{align*}
    Therefore
    \begin{align*}
    \ep^{(1)} 
    & = \frac{\sqrt{r_M+r_1}  \kappa  }{\lam_{r_M + r_1}(P) }  \\
    & = O\bigg(\frac{\sqrt{r_M+r_1}\bigg \|\|A - P\| \bigg \|_{\psi_1}  }{\lam_{r_M + r_1}(P) } \bigg)  \\
    & = O\bigg( \frac{\sqrt{r_M+r_1} \sqrt{\rho_n n}  }{\lam_{r_M + r_1}(P) } \bigg)  \\
     & = O \bigg( \frac{1 }{  \sqrt{\rho_n n}} \bigg) 
    \end{align*}
    and 
    \begin{align*}
    \ep^{(2)} 
    & = \frac{\tau \sqrt{n(r_M+r_2) (\sig_{r_M + r_2}^2(W) + p ) }}{\sig_{r_M + r_2}^2(W)} \wedge \sqrt{r_M+r_2} \\
    & =  \bigg[ \tau \bigg( \frac{n(r_M+r_2)}{\sig_{r_M + r_2}^2(W)}  + \frac{n(r_M+r_2)p}{\sig_{r_M + r_2}^4(W)} \bigg)^{1/2} \bigg] \wedge \sqrt{r_M+r_2} \\
     & = O \bigg( \bigg[ \tau \bigg( \frac{1 }{p } + \frac{1}{np}\bigg)^{1/2} \bigg]  \wedge 1 \bigg ) \\
     & = O \bigg( \frac{\tau}{\sqrt{p}} \bigg).
    \end{align*}
    Hence $\ep^{(1)} + \ep^{(2)} = O\bigg( \frac{1 }{ \sqrt{\rho_n n}}  + \frac{\tau }{ \sqrt{p}} \bigg)$, so that 
    \begin{align*}
    \E[ d(\wh{M}, M)] 
        & = O\bigg(\frac{\sqrt{r_M}}{\del}[\ep^{(1)} +  \ep^{(2)}] \bigg) \\
        & = O\bigg( \frac{1 }{  \sqrt{\rho_n n}}  + \frac{\tau }{ \sqrt{p}} \bigg) 
    \end{align*}
    and 
    \begin{align*}
    \E[ d(\wh{R}^{(k)}, R^{(k)})] 
        & = O\bigg(\frac{\sqrt{r_M r_k}}{\del}[\ep^{(1)} +  \ep^{(2)}] \bigg) \\
        & = O \bigg( \frac{1 }{  \sqrt{\rho_n n}}  + \frac{\tau }{ \sqrt{p}} \bigg).
    \end{align*}
    Since $\rho_n n = \om (\log n)$, we have that $(\rho_n n)^{-1/2} = o((\log n)^{-1/2})$ and consistency follows. 
    \end{proof}

	\bibliographystyle{chicago}
	\bibliography{References}
\end{document}